\documentclass[conference]{IEEEtran}

\newif\ifdraft
\drafttrue

\newif\iftechreport
\techreporttrue

\usepackage{tikz}
\usepackage{pgfplots}
\usepackage{color}
\usepackage{xspace}
\usepackage{amsfonts}
\usepackage{amsmath}
\usepackage{amsthm}
\usepackage{algorithm}
\usepackage{algorithmicx}
\usepackage{todonotes}
\usepackage[hidelinks]{hyperref}
\PassOptionsToPackage{hidelinks}{hyperref}
\usepackage[capitalise,noabbrev]{cleveref}
\usepackage{mdframed}
\usepackage{paralist} %
\usepackage[inline]{enumitem}
\usepackage[noend]{algpseudocode}
\usepackage[export]{adjustbox}
\usepackage{cite}
\usepackage{wrapfig}
\PassOptionsToPackage{sort}{cite}

\ifdraft
\fi

\renewcommand{\paragraph}[1]{\smallskip \noindent \textbf{#1.}~}

\newcommand{\vpcn}[0]{VPCN\xspace}
\newcommand{\uid}[0]{\ensuremath{v}}
\newcommand{\function}[1]{\ensuremath{\textit{#1}}}
\newcommand{\userid}{\uid}
\newcommand{\TT}{\mathcal{T}}
\newcommand{\BB}{\mathcal{B}}
\newcommand{\PP}{\mathcal{P}}

\newcommand{\union}[0]{\ensuremath{\cup}\xspace}
\newcommand{\set}[1]{\ensuremath{\{#1\}}\xspace}

\newcommand{\GG}[0]{\ensuremath{\mathcal{G}}\xspace}
\newcommand{\VV}[0]{\ensuremath{\mathcal{V}}\xspace}
\newcommand{\EE}[0]{\ensuremath{\mathcal{E}}\xspace}
\newcommand{\EEpay}[0]{\ensuremath{\EE_p}\xspace}
\newcommand{\EEvirtual}[0]{\ensuremath{\EE_v}\xspace}
\newcommand{\blockchain}[0]{\ensuremath{\mathbb{B}}\xspace}
\newcommand{\XX}{\mathcal{X}}

\newcommand{\pcid}[2]{\ensuremath{\textit{pc}_{\langle#1,#2\rangle}}\xspace}
\newcommand{\vcid}[2]{\ensuremath{\textit{vc}_{\langle#1,#2\rangle}}\xspace}
\newcommand{\cid}[2]{\ensuremath{\textit{c}_{\langle#1,#2\rangle}}\xspace}
\newcommand{\bal}[0]{\ensuremath{\beta}\xspace}
\newcommand{\balanceonchain}[0]{\ensuremath{\beta^{{\small \textit{on-chain}}}}\xspace}

\newcommand{\fee}[0]{\ensuremath{f}\xspace}
\newcommand{\feeprop}[0]{\ensuremath{p}\xspace}
\newcommand{\feecreate}[0]{\ensuremath{f_e}\xspace}

\newcommand{\openpc}[0]{\function{openPC}\xspace}
\newcommand{\closepc}[0]{\function{closePC}\xspace}
\newcommand{\openvc}[0]{\function{openVC}\xspace}
\newcommand{\closevc}[0]{\function{closeVC}\xspace}
\newcommand{\updatepvc}[0]{\function{update\{P,V\}C}\xspace}
\newcommand{\pay}[0]{\function{pay}\xspace}

\newcommand{\lukas}[1]{#1}

\newcommand{\bigO}{\mathcal{O}}

\newcommand{\amount}{\ensuremath{\alpha}}
\newcommand{\capacityLocked}{\ensuremath{capacity\_locked}}
\newcommand{\advBudget}{\ensuremath{adv\_budget}}
\newcommand{\totalNetworkCapacity}{\ensuremath{total\_network\_capacity}}
\newcommand{\costBenefit}{\ensuremath{cost\_benefit}}
\newcommand{\occ}{\ensuremath{occ}}
\newcommand{\numPay}{\ensuremath{num\_pay}}
\newcommand{\rPay}{\ensuremath{r\_pay}}
\newcommand{\val}{\ensuremath{val}}
\newcommand{\routeVc}{\ensuremath{route\_vc}}
\newcommand{\establishVc}{\ensuremath{establish\_vc}}
\newcommand{\routePcn}{\ensuremath{route\_pcn}}
\newcommand{\feeRatio}{\ensuremath{fee\_ratio}}

\usepackage[utf8]{inputenc}
\usepackage[T1]{fontenc}
\usepackage{varwidth}
\usepackage{xcolor}
\usetikzlibrary{positioning, shapes.geometric, shapes, calc}

\tikzset{
    switch/.style={
        circle, 
        draw=black!70, 
        fill=white,  
        thick, 
        minimum size=0.7cm
    },
    txt/.style={
        draw=none,
        rectangle
    },
    legb/.style={
        draw=SkyBlue,
        fill=SkyBlue,
        rectangle
    },
    legg/.style={
        draw=LimeGreen,
        fill=LimeGreen,
        rectangle
    }
}

\mdfsetup{frametitlealignment=\center, innerleftmargin=5pt, innerrightmargin=3pt}
\newcommand{\mdframeSkip}{15pt}

\newcommand{\pseudocomment}[1]{\textcolor{gray}{//#1}\xspace}

\newcommand*\Let[2]{\State #1 $\gets$ #2}
\algrenewcommand\algorithmicrequire{\textbf{Precondition:}}
\algrenewcommand\algorithmicensure{\textbf{Postcondition:}}

\newcounter{concounter}
\newcommand{\newcon}{\stepcounter{concounter}\textbf{(C\theconcounter)}~}

\theoremstyle{definition}
\newtheorem{definition}{Definition}
\newtheorem{theorem}{Theorem}

\begin{document}

\title{Optimizing Virtual Payment Channel Establishment in the Face of On-Path Adversaries}

\date{}

\author{Lukas Aumayr\textsuperscript{1,5} \quad Esra Ceylan\textsuperscript{2} \quad Yannik Kopyciok\textsuperscript{3} \quad Matteo Maffei\textsuperscript{1,5} 
\\ Pedro Moreno-Sanchez\textsuperscript{4} \quad Iosif Salem\textsuperscript{3} \quad Stefan Schmid\textsuperscript{3}\\
{\small \textsuperscript{1}TU Wien 
\quad
\textsuperscript{2}University of Vienna 
\quad
\textsuperscript{3}TU Berlin
\quad
\textsuperscript{4}IMDEA Software Institute
\quad}\\{\small
\textsuperscript{5}Christian Doppler Laboratory Blockchain Technologies for the Internet of Things
}
}

\IEEEoverridecommandlockouts
\IEEEpubid{\begin{minipage}{\columnwidth}\ \\[12pt]
 This is the extended technical report of the work accepted at IFIP Networking 2024.
\end{minipage}
\hspace{\columnsep}\makebox[\columnwidth]{ }}

\pagestyle{plain}

\maketitle

\begin{abstract}
Payment channel networks (PCNs) are among the most promising solutions to the scalability issues in permissionless blockchains, by allowing parties to 
pay each other off-chain through a path of payment channels (PCs). 
However, routing transactions comes at a cost which is proportional to the number of intermediaries, since 
each charges a fee for the routing service.
Furthermore, analogous to other networks, malicious intermediaries in the 
payment path can lead to security and privacy threats. 
Virtual channels (VCs), i.e., bridges over PC paths, 
mitigate the above PCN issues, as an intermediary participates only 
once to set up the VC and is then excluded from every future VC transaction. 
However, similar to PCs, creating a VC has a cost that must be paid out of 
the bridged PCs' balance. 
Currently, we are missing guidelines to where and how many VCs to set up. 
Ideally, VCs should minimize transaction costs while mitigating security and privacy threats from on-path adversaries.

In this work, we address for the first time  the VC setup problem, formalizing it as an optimization problem.
We present an integer linear program (ILP) to compute the globally optimal VC setup strategy in terms of transaction costs, security, and privacy. 
We then accompany the computationally heavy ILP with a fast local greedy algorithm. Our model and algorithms can be used with any on-path adversary, given that its strategy can be expressed as a set of corrupted nodes that is estimated by the honest nodes. We conduct an evaluation of the greedy algorithm over a snapshot of the Lightning Network (LN), the largest Bitcoin-based PCN. Our results confirm on real-world data that our greedy strategy minimizes costs while 
protecting against security and privacy threats of on-path adversaries.
These findings may serve the LN community as guidelines for the  deployment of VCs.

\end{abstract}


\section{Introduction}

Permissionless cryptocurrencies face severe scalability challenges, 
as they rely on a set of mutually untrusted users located across the world to 
maintain a distributed and publicly verifiable transaction ledger. 
The transaction throughput today is limited 
to tens of transactions per second at best, while
transactions can take up to 60 minutes to be confirmed.

Payment channels (PC) have emerged as one of the most promising scalability solutions, and instances 
such as the Lightning Network~\cite{poonBitcoinLightningNetwork} are gaining traction. 
In this approach, 
Alice and Bob can create a PC between them with a single on-chain 
transaction that transfers their coins into an escrow (or multi-signature) 
controlled by both of them with the additional guarantee that they can get refunded at a mutually agreed-upon time. 
After that, 
Alice and Bob can pay each other off-chain by exchanging authenticated 
copies of the updated balances in the escrow. 
Finally, the PC is closed with 
an on-chain transaction representing  
the last authenticated distribution of coins.

PCs can be linked to form a network, also called a payment channel network (PCN), where any two users can perform a payment if they are connected by a path of PCs. The payment in the PC between Alice and the first intermediary is forwarded along the intermediary PCs until it reaches Bob. A key challenge in 
this approach is then to ensure that the balance updates of all PCs in the path are 
atomic to prevent any intermediary (i.e., Ingrid) from trivially stealing the money 
by denying forwarding it.

State-of-the-art techniques to construct atomic multi-hop payments~\cite{malavoltaAnonymousMultiHopLocks2019,aumayrBlitzSecureMultiHop2021,malavolta2017concurrency,heilman2017tumblebit,tairiMbox2AnonymousAtomic2019,eggerAtomicMultiChannelUpdates2019,jourenkoPaymentTreesLow2020} 
require that intermediaries are involved in every single payment. This approach 
brings several disadvantages: (i) reduction of the payment reliability (e.g., 
Ingrid may simply be offline or crash); (ii) increase in the payment latency since  
additional PCs are required; (iii) high payment costs as 
each intermediary charges a fee per transaction for providing the routing service; 
and (iv) possible leakage of sensitive information to the intermediaries, which opens the door to a number of security and privacy issues, such as route hijacking~\cite{aft20}, wormhole attacks~\cite{malavoltaAnonymousMultiHopLocks2019} or user anonymity~\cite{malavolta2017concurrency}, just to name a few.

Recently, the concept of virtual channels (VCs)~\cite{dziembowski2017perun,aumayrBitcoinCompatibleVirtualChannels,DBLP:conf/cans/JourenkoLT20,cryptoeprint:2021:855} has been proposed to improve 
upon the aforementioned drawbacks of PCs. A VC can be seen as a bridge over two PCs. 
For instance, assume that Alice and Bob have a PC with 
an intermediary, Ingrid. In order to set up a VC, Ingrid must collaborate and coordinate 
with Alice and Bob to lock coins in their corresponding PCs in order to 
use those coins 
to build a VC directly between Alice and Bob. This approach brings 
the following benefits: (i) Alice and Bob can pay each other ``as if they had a PC between them'', that is, without the involvement of Ingrid;
 (ii) payment latency is reduced to one hop; 
 (iii)  payment fees charged by the intermediaries for their routing service are avoided;
 and (iv)  the details of every single payment are not revealed to possibly malicious or curious intermediaries.
Note that intermediaries still charge fees for the coordination service when establishing the virtual channel.

A crucial question, not yet addressed in the literature,  is \emph{what strategy should users follow to open VCs while optimizing the cost-effectiveness, as well as on-path security and privacy benefits provided by VC networks?} 
This is an optimization problem, given that the funding to be locked, and thus the number of VCs a party can establish, is limited by the number of underlying PCs and the amount of coins that are locked on them.
To ensure \textit{on-path} security and privacy we provide a modular framework for preventing attacks. Our algorithms use as input a set of nodes that the honest nodes estimate to be corrupted. The honest nodes derive the set of potentially corrupted nodes based on their assumptions on the adversary. We demonstrate how our framework functions through three exemplary well-studied attacks in the literature.
Note that, while several existing works have studied from a game theoretic perspective how a  PCN should evolve based on the fee optimization goal of the  users~\cite{DBLP:conf/esorics/AvarikiotiSW19,DBLP:conf/fc/AvarikiotiH0W20,DBLP:conf/fc/ErsoyRE20}, none of them considered virtual channels, nor on-path security and privacy goals.  %

We make the following \textbf{\emph{contributions}}. First, we address the VC setup problem, formalizing it as 
an optimization problem of three distinct goals: (i) cost-effectiveness of the transactions (i.e., fees)  while providing (ii) security and (iii) privacy guarantees against on-path adversaries, and prove that the optimization problem is NP-hard. On-path adversaries account for a significant 
share of attacks in the PCN-related literature: e.g., they  may aim to perform denial-of-service and wormhole attacks, or to harm value privacy and relationship anonymity properties, among many others~\cite{tang2020privacy,icissp20,malavolta2017concurrency,tripathy2020mappcn,aft20,spam,malavoltaAnonymousMultiHopLocks2019}. Such attacks have been shown to potentially have a severe impact in practice~\cite{tikhomirov2020quantitative}. On-path adversaries can do damage depending on the attack that they are carrying out. In this work, we provide a general framework for mitigating attacks of on-path adversaries and study three exemplary attacks. 
Specifically, our algorithms use as input the set of nodes which the honest nodes estimate to be corrupted, given their assumptions on the adversarial strategy. The derivation of this set by the honest nodes is orthogonal to our solutions. To demonstrate our solutions we utilize adversarial strategies that affect the largest fraction of payments \cite{tikhomirov2020quantitative,kapposEmpiricalAnalysisPrivacy2021,DBLP:journals/corr/abs-2103-08576,10.1145/3465481.3465761} and focus on value privacy, relationship anonymity, and the wormhole attack. \lukas{Note that the adversarial strategy can be replaced easily by any other strategy.}

Second, we analytically show a synergy between the different VC optimization objectives. In particular, we prove that minimizing transaction fees by the appropriate use of VCs also prevents attacks from on-path adversaries,  such as those against 
value privacy and relationship anonymity, or wormhole attacks. 
In practice, this implies that users can set up 
their VCs following a single strategy to minimize their transaction costs, and as a side 
benefit, they will be secure against on-path adversaries. We demonstrate the latter for the three exemplary on-path attacks on security and privacy in study.

Third, and motivated by the uncovered synergy between the objectives, we describe   
concrete approaches to devise fee optimization strategies which mitigate on-path security and privacy attacks (and specifically value privacy, relationship anonymity, and wormhole attacks). 
In particular, we present both an efficient approach (based on a greedy routing algorithm) to optimize the cost-effectiveness, security, and privacy of PCNs using VCs, and a rigorous and exact approach based on integer linear programming (ILP), which is computationally intractable (we also propose how to reduce the running time of the ILP). \lukas{The network topology of PCNs such as the Lightning Network is known publicly. In our exact ILP-based approach, we additionally assume that all transactions we want to route are known globally, in order to find the globally optimal solution. Our greedy algorithm, on the other hand, can be applied locally, using only the information of individual nodes.}

Finally, we evaluate our greedy optimization approach on a recent snapshot of the Lightning Network (LN). We show that our transaction cost minimization strategy is efficient and effective, and indeed subsumes the strategies 
to optimize for on-path security and privacy. We find that depending on how many payments two endpoints plan to conduct via the virtual channel, the routing cost can be reduced significantly, for example, to about half compared to a normal payment for two consecutive payments,
or to about 3\% for 50 consecutive payments. In addition to this cost reduction, other users can utilize these virtual channels to route their payments through a potentially cheaper path. 

To summarize, for the first time, we present both an analytical and an empirical study of the impact of using VCs in (current) PCNs in terms of cost-effectiveness of the transactions as well as security and privacy guarantees. 
The results of this work motivate the deployment of VCs and we hope that they can encourage the PCN community and developers to include VCs within current PCNs software and make them accessible to the PCN users.

\paragraph{Paper organization} We introduce background knowledge on PCNs in Section \ref{sec:pcns-background} and present our model and problem formulation in Section \ref{sec:model}.
We present our algorithms in sections \ref{sec:ILPfull} (exact) and \ref{sec:greedy} (greedy), and evaluate the greedy algorithm in Section \ref{sec:heuristics}.

\section{Background and problem overview}\label{sec:background}

\paragraph{Payment channel networks (PCNs)}
\label{sec:pcns-background}
A PCN~\cite{malavolta2017concurrency} is a directed 
graph $\GG := (\VV, \EE)$. Nodes $\VV$ represent users and 
edges $\{e_{i,j}, e_{j,i}\} \subset \EE$ represent PCs between users. 
The weight on a directed edge denotes the amount of remaining 
coins that can be forwarded on that direction. 
For every pair of edges $\{e_{i,j}, e_{j,i}\}$, users $\userid_i$ and $\userid_j$ can exchange any part of their balance freely.
Moreover, each directed edge $e_{i,j}$ between users $\userid_i$ and $\userid_j$ is associated with two  
non-negative numbers, the base fee $f_i$, and the proportional fee $p_i$, that together determine the fees that each user charges for forwarding the payments. 
For a forwarded amount $\alpha$ via $e_{i,j}$, $\userid_i$ charges $\textit{fee}(e_{i,j}, \userid_i) = f_i + p_i\cdot \alpha$.
We denote a PC $\{e_{i,j}, e_{j,i}\}$ with the tuple $(\pcid{\uid_i}{\uid_j}, \bal_i, \bal_j, \fee_i,$ $\fee_j, \feeprop_i, \feeprop_j)$, where $\beta_{k\in\{i,j\}}$ is the initial balance of each node upon channel creation.

The success of a payment between two users depends on the capacity available
in the path connecting the sender $s$ to the receiver $r$. Assume that 
$s$ wants to pay $\alpha$ coins to $r$ and that they are connected through 
a path $s \rightarrow u_1 \rightarrow \ldots \rightarrow u_n \rightarrow r $. 
The fees charged for every node in the path depend on the forwarded amount.
That is, $\userid_n$ charges $fee_n = f_{n,n+1} + p_{n,n+1}\cdot \alpha$ and in general  $\userid_{j}$ charges $fee_{j} = f_{j,j+1} + p_{j,j+1} \cdot (\alpha + \sum_{k=j+1}^n fee_k)$, for $j=1,\ldots,n$ (each node forwards $\alpha$ and the forwarding fees of the remaining nodes in the path).
Such a payment is successful if 
(i) $s$ starts the payment with a value $\alpha^* := \alpha + \sum_{j=1}^{n} fee_j$ and
(ii) every edge on the path has a balance 
of at least $\alpha'_i$, where $\alpha'_i := \alpha^* - \sum_{j=1}^{i-1} fee_j$ (the initial payment value $\alpha$ minus the fees charged 
by the previous users in the path), $e_{j,j+1} = (u_j, u_{j+1})$, and $u_{n+1}=r$. 
If the payment is successful, the balance of
every edge $e_{j,j+1}$ on the path from $s$ to $r$ is decreased by $\alpha'_i$, while the balance of every edge $e_{j+1,j}$ is increased by $\alpha'_i$.

\begin{figure}
    \centering
    \includegraphics[scale=0.56]{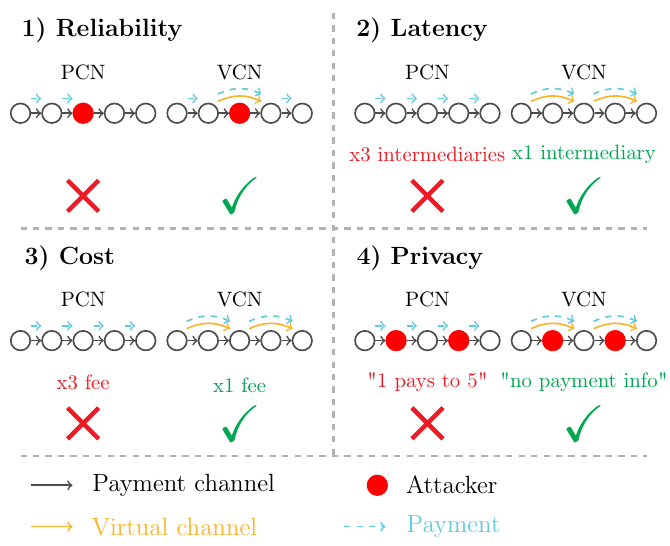}
    \vspace{-0.8em}
    \caption{Comparison between PCN and VCN.}
    \label{fig:comparison-pcn-vcn}
\end{figure}
\paragraph{PCN challenges} 
For successful payments, intermediaries must actively participate and must not disturb them, either 
actively (e.g., dropping it) or passively (e.g., being offline). Thus, PCN payments suffer from the following drawbacks:

\textit{Reliability:} If intermediaries are offline or do not forward the payment (e.g., the red user in \cref{fig:comparison-pcn-vcn}), the payment fails.

\textit{Latency:} The time to process a payment is directly proportional 
to the number of intermediate users. 
E.g., the latency of the payment 
shown in \cref{fig:comparison-pcn-vcn} (latency section) 
could be reduced if a shorter path between nodes $1$ and $5$  existed. 

\textit{Cost:} The payment cost is 
 proportional to the number of intermediate users, since each 
charges a routing fee.

\textit{Privacy:} Each intermediary learns  
sensitive information. Recent work \cite{malavolta2017concurrency,greenBoltAnonymousPayment2017} 
has shown that intermediaries can learn details about who pays what to whom in the 
currently deployed Lightning Network. While alternative payment mechanisms that hide (some of) the information 
required in such payment exist, e.g.,~\cite{malavoltaAnonymousMultiHopLocks2019,aumayrBlitzSecureMultiHop2021}, they have not been adopted yet and still protect only some sensitive information but not other (e.g., the payment amount) and also do not decrease routing fees.

\paragraph{Virtual channels (VCs)}
\label{sec:vc-background}
Bypassing intermediaries can mitigate these drawbacks. One could build a new PC, but this requires an expensive on-chain transaction and additional funds.
Instead, a VC can be created off-chain between two users, say Alice and Bob, 
who have a PC with a common intermediary, say Ingrid. 
Using a 3-party protocol, the users can block coins in the underlying PCs 
and move them into the VC between Alice and Bob. 
After that,  
Alice and Bob can perform arbitrarily many payments without involving Ingrid. The amount of VCs that 
can be created are thus limited by the balances of the underlying 
PCs. Yet, it is interesting to deploy VCs 
as they provide several advantages over PCNs.

\textit{Reliability:} Payments are carried out without involving the intermediary user, who cannot thus disturb it either 
actively (e.g., dropping it) or passively (e.g., being offline). 
In \cref{fig:comparison-pcn-vcn}, the malicious node $3$ does not participate 
in the payment between $1$ and $5$ as it is omitted by the VC between $2$ and $4$.

\textit{Latency:} VCs lead to shorter paths. 
Since there are fewer intermediate users, the latency of the overall payment is reduced. In the 
running example, the latency is reduced from 3 to 1 intermediaries, assuming that 
two VCs have been created. 

\textit{Cost:} Assume, for simplicity, that users charge the same fees for forwarding a 
payment through a PC and a VC. In such a case, 
as with latency, the fact that VCs lead to shorter paths, 
can also help to reduce the overall payment cost in terms of fees.
In \cref{fig:comparison-pcn-vcn}, the transaction cost using VCs 
is reduced to the fee charged by the only intermediary that is involved, avoiding thus 
the fees charged by nodes 2 and 4.

\textit{Privacy:} The fact that fewer intermediaries are 
participating in the payment improves the privacy of the overall payment. 
And although intermediaries are part of the 3-party creation of the VC 
and thus learn who are the two VC endpoints, they no longer see the amounts of the individual 
payments routed through the VC. 
For instance, in Figure \ref{fig:comparison-pcn-vcn}, the malicious node $2$ would learn that 
there exists a VC between nodes $1$ and $3$ as it needs to help them to set the 
channel up, but afterwards the node $2$ does not learn when a VC is used. 

\paragraph{VCs in practice}
Despite the advantages provided by VCs, 
we currently lack a comprehensive analysis leading to a set of guidelines to help 
the users decide when to open VCs, with what neighbors, and under what 
circumstances. Ideally, a user would like to open a VC with 
every other user in the network. Unfortunately, this is not possible since each user 
has a limited budget, i.e., the amount of coins available on her PCs which need to be locked to create a VC. In this state of affairs, the following questions arise: 
how should a user choose which neighbor to open a VC with? 
how many payments are required to amortize the cost of opening a VC? 
what strategy should a user follow to maximize the security and privacy gains against on-path adversaries 
when opening VCs?

\section{Modelling virtual payment channel networks}
\label{sec:model}

We introduce a more formal model of virtual payment channel networks (VPCNs). We will then discuss the security and privacy threats by on-path adversaries, define the studied optimization goal on VPCNs, and show its NP-hardness.

\begin{figure}[b!]
\centering
\begin{footnotesize}  
\resizebox{\columnwidth}{!}{
\hspace{-2mm}
\begin{minipage}{\columnwidth}
    \footnotesize 
\begin{tabular}{cc}
    \hspace{-3mm}
    \fbox{\parbox{0.52965\textwidth}{
    \footnotesize
	\underline{\openpc$(\uid_1, \uid_2, \bal_1, \bal_2, \fee_1, \fee_2, \feeprop_1, \feeprop_2)$}
	    \begin{tabular}{rl}
	       $\uid_{i \in \set{1,2}}$: &  Nodes  \\
	       $\bal_{i \in \set{1,2}}$: & PC initial capacity \\
	       $\fee_{i \in \set{1,2}}$: & Base routing fees \\
	       $\feeprop_{i \in \set{1,2}}$: & Proportional routing fee
	    \end{tabular}
	    
	    \smallskip
			\begin{asparaitem}
				\item If $\blockchain[\uid_1] < \bal_1$ or  $\blockchain[\uid_2] < \bal_2$, abort. Else, create a new payment channel $(\pcid{\uid_1}{\uid_2}, \bal_1, \bal_2, \fee_1, \fee_2,  \feeprop_1, \feeprop_2) \in \EEpay$
		 	    \item Update blockchain as $\blockchain[\uid_1] = \blockchain[\uid_1] - \bal_1$ and $\blockchain[\uid_2] = \blockchain[\uid_2] - \bal_2$
		 \end{asparaitem}
		 \vspace{0.4mm}
	}}

&
    \hspace{-5.4mm}
    \fbox{
    \parbox{0.425\textwidth}{
    \footnotesize
	\underline{\closepc$(\pcid{\uid_1}{\uid_2})$}
	
	\begin{tabular}{r l }
	   $\pcid{\uid_1}{\uid_2}$:  &  PC identifier
	\end{tabular}
	
	\smallskip 
	
	\begin{asparaitem}
	    \item Let $(\pcid{\uid_1}{\uid_2}, \bal_1, \bal_2,\allowbreak \fee_1,\allowbreak \fee_2, \feeprop_1, \feeprop_2)$ be the corresponding entry in $\EEpay$. If such entry does not exist, abort.
	   	\item Set $\blockchain[\uid_1] = \blockchain[\uid_1] + \bal_1$, $\blockchain[\uid_2] = \blockchain[\uid_2] + \bal_2$ and remove from $\EEpay$:  $(\pcid{\uid_1}{\uid_2}, \bal_1, \bal_2,\allowbreak \fee_1,\allowbreak \fee_2,\allowbreak \feeprop_1, \feeprop_2)$ 
	\end{asparaitem}
	}}
\end{tabular}

    \vspace{-0.29mm}
    \begin{tabular}{cc}
    \hspace{-3mm}
    \fbox{\parbox{0.67\textwidth}{
    \footnotesize
	\underline{\scriptsize\openvc$(\cid{\uid_1}{\uid_I}, \cid{\uid_I}{\uid_2},\bal_1, \bal_2, \fee_1, \fee_2, \feeprop_1, \feeprop_2, \feecreate)$}\\
	\begin{tabular}{r l}
	   $\cid{\uid_1}{\uid_I}$, $\cid{\uid_I}{\uid_2}$:  & PC or VC identifiers \\
	    $\bal_{i \in \set{1,2}}$: & Initial VC balance \\
	    $\fee_{i \in \set{1,2}}$: & Base routing fee \\
	    $\feeprop_{i \in \set{1,2}}$: & Proportional r. fee \\
	    $\feecreate$: & Establishing fee
	\end{tabular}
	
	\smallskip 
	
	\begin{asparaitem}
		\item Let $(\cid{\uid_1}{\uid_I}, \bal'_1, \bal'_I, \fee'_1, \fee'_I, \feeprop'_1, \feeprop'_I)$ and $(\cid{\uid_I}{\uid_2},\allowbreak \bal''_I,\allowbreak \bal''_2,\allowbreak \fee''_I,\allowbreak \fee''_2,\allowbreak \feeprop''_I,\allowbreak \feeprop''_2)$ be the entries in $\EEpay$ or $\EEvirtual$ corresponding to $\cid{\uid_1}{\uid_I}$ and $\cid{\uid_I}{\uid_2}$. If $(\bal_1 + \feecreate) > \bal'_1$ or $\bal_2 > \bal'_I$ or  $\bal_2 > \bal''_2$ or $\bal_1 > \bal''_I$, abort.
		\item Update the entries in $\EEpay$ or $\EEvirtual$ as $(\cid{\uid_1}{\uid_I}, \bal'_1 - (\bal_1 + \feecreate), \bal'_I - \bal_2 + \feecreate, \fee'_1, \fee'_I, \feeprop'_1, \feeprop'_I)$ and  $(\cid{\uid_I}{\uid_2}, \bal''_I - \bal_1, \bal''_2 - \bal_2, \fee''_I, \fee''_2, \feeprop''_I, \feeprop''_2)$
		\item Add  $(\vcid{\uid_1}{\uid_2}, \bal_1, \bal_2, \fee_1, \fee_2, \feeprop_1, \feeprop_2)$ in $\EEvirtual$
	\end{asparaitem}
	\vspace{-1.1mm}
	}}
		&

    \hspace{-5.4mm}
    \fbox{\parbox{0.2952\textwidth}{
    \footnotesize
	\underline{\closevc$(\vcid{\uid_1}{\uid_2})$ }

	\begin{tabular}{r l}
	   \hspace{-2mm}$\vcid{\uid_1}{\uid_2}$:  &  \hspace{-4mm}VC identifier 
	\end{tabular}
	
	\smallskip
	
	\begin{asparaitem}
	    \item In $\EEvirtual$, remove the corresponding entry $(\vcid{\uid_1}{\uid_2},\allowbreak \bal_1,\allowbreak \bal_2,\allowbreak \fee_1,\allowbreak \fee_2,\allowbreak \feeprop_1,\allowbreak \feeprop_2)$ if it exists. If such entry does not exist, abort.
	    
	    \item Update entries in $\EEpay$ as $(\pcid{\uid_1}{\uid_I}, \bal'_1 + \bal_1,\allowbreak \bal'_I + \bal_2, \fee'_1, \fee'_I, \allowbreak\feeprop'_1,\allowbreak \feeprop'_I)$ and $(\pcid{\uid_I}{\uid_2},\allowbreak \bal''_I\allowbreak + \bal_1,\allowbreak \bal''_2\allowbreak + \bal_2,\allowbreak \fee''_I,\allowbreak \fee''_2,\allowbreak \feeprop''_I,\allowbreak \feeprop''_2)$
		 	
	\end{asparaitem}
        \vspace{2.4mm}
	}}
    \end{tabular}

    \vspace{-0.25mm}
    \hspace{-0.55mm}
    \fbox{\parbox{0.99\textwidth}{
	\underline{\updatepvc$(\cid{\uid_1}{\uid_2}, \bal)$}
    
    \begin{tabular}{r l}
       $\cid{\uid_1}{\uid_2}$:  & Channel identifier 
    \end{tabular}
    
    \smallskip
    
    \begin{asparaitem}
    \item Let 
	$(\cid{\uid_1}{\uid_2}, \bal_1, \bal_2, \fee_1, \fee_2, \feeprop_1, \feeprop_2)$ be the corresponding entry in $\EEpay \union \EEvirtual$. If such entry does not exist or it exists but $\bal_1 < \bal$, abort.
    
    \item Update the %
    channel as follows $(\cid{\uid_1}{\uid_2}, \bal_1 - \bal, \bal_2 + \bal, \fee_1, \fee_2, \feeprop_1, \feeprop_2)$
    \end{asparaitem}
    }}

    \vspace{-0.3mm}
    \hspace{-0.55mm}
    \fbox{\parbox{0.99\textwidth}{
	 \underline{\pay$((\cid{s}{\uid_1}, \ldots, \cid{\uid_n}{r}), \bal)$}
    
        \begin{tabular}{r l}
           $(\cid{s}{\uid_1}, \ldots, \cid{\uid_n}{r})$:  & List of channels \\
            $\bal$: & Payment amount 
        \end{tabular}
        
        \smallskip
        
        \begin{asparaitem}
            \item If the channels do not form a path from sender $s$ to receiver $r$, abort.

            \item %
                If there is a channel $(\cid{\uid_i}{\uid_{i+1}},$ $\bal_i, \bal_{i+1}, \fee_i, \fee_{i+1}, \feeprop_{i}, \feeprop_{i+1})$, $i = 0,1,\ldots, n$ ($s=v_0$ and $r=v_{n+1}$), for which $\beta_i < send_i$, abort.%

		 	\item %
		 	Update each channel in the path: $(\cid{\uid_i}{\uid_{i+1}}, \bal_i - send_i, \bal_{i+1} + send_i, \fee_i, \fee_{i+1}, \feeprop_{i}, \feeprop_{i+1})$, $i=0,1,\ldots, n$.%
        \end{asparaitem}
        
        Recursive definitions for a payment path $(s=v_0,v_1, \ldots, v_n, r=v_{n+1})$.\\
		 $send_\ell$, $\ell = 0,\ldots, n$, is the amount that node $v_\ell$ sends to $v_{\ell+1}$:
		 $send_\ell = \beta + \sum_{i=\ell+1}^n fee_i$.\\
		 $fee_\ell$, $\ell = 1,\ldots, n$, is the amount that node $v_\ell$ charges (keeps) for forwarding the payment
		 $fee_\ell = f_\ell + p_\ell\cdot send_\ell =$ $f_\ell + p_\ell(\beta + \sum_{i=\ell+1}^n fee_i)$.		 
	    We say a sum starting from a higher index than its ending index equals zero.
    }}

\end{minipage}
}
\vspace{-0.2cm}
\caption{Operations in a VPCN. %
$v_1$ and $v_2$ share the VC establishing fee $f_e$.
} \label{sec:opvpcn}
\end{footnotesize}
\end{figure}

\begin{definition}[\vpcn]
\label{def:vpcn}
A virtual payment channel network, \vpcn, is defined as a graph $\GG := (\VV, \EE)$ where $\VV$ denotes the set of users in the network and $\EE := \EEpay \union \EEvirtual$ denotes the set of channels. In particular, $\EEpay$ denotes the set of payment channels and $\EEvirtual$ denotes the set of VCs. 
Each payment channel is defined by a tuple $(\pcid{\uid_1}{\uid_2}, \bal_1, \bal_2, \fee_1,$ $\fee_2,  \feeprop_1, \feeprop_2)$\footnote{In~\cite{malavolta2017concurrency}, a PC contains a timeout parameter. This is no longer required~\cite{poonBitcoinLightningNetwork}.}, where $\pcid{\uid_1}{\uid_2}$ denotes a payment channel identifier, $\bal_{i \in \set{1,2}}$ denotes the current balance of the node $\uid_{i \in \set{1,2}}$, $\fee_{i \in \set{1,2}}$ is the base fee and $\feeprop_{i \in \set{1,2}}$ the fee rate (proportional to the amount paid) charged to use this channel in each direction, respectively.
Analogously, a VC is defined by a tuple $(\vcid{\uid_1}{\uid_2}, \bal_1, \bal_2,$ $\fee_1, \fee_2,  \feeprop_1, \feeprop_2, \feecreate)$, where $\feecreate$ denotes the VC establishment fee.%

A \vpcn is defined with respect to a blockchain $\blockchain$ that stores publicly accessible entries of the form $(\uid, \balanceonchain)$ where $\uid$ denotes an address of the underlying blockchain and $\balanceonchain$ denotes its on-chain balance. 
For readability, we hereby use $\blockchain[\uid]$ to denote the on-chain balance of $\uid$ in $\blockchain$. A \vpcn exposes the operations expressed in~\autoref{sec:opvpcn}.  

\end{definition}

\paragraph{Security and privacy for on-path adversaries}
On-path adversaries may cause a diverse set of attacks in PCNs~\cite{tang2020privacy,icissp20,malavolta2017concurrency,tripathy2020mappcn,aft20,spam,malavoltaAnonymousMultiHopLocks2019} and with significant impact~\cite{tikhomirov2020quantitative}. 
We employ VCs to defend against on-path adversaries, by bypassing corrupted nodes (cf. Figure~\ref{fig:comparison-pcn-vcn}). We chose to investigate three representative attacks: value privacy, relationship anonymity, and wormhole attacks \cite{malavolta2017concurrency,malavoltaAnonymousMultiHopLocks2019}. %
We chose these attacks because they are well studied in the literature and note that our approach directly generalizes to other on-path adversarial attacks, such as denial-of-service attacks~\cite{aft20}, which we later show in \cref{sec:strategy_relation}.

\paragraph{Optimal adversarial strategy} We assume that the adversary has a budget $\BB$ for corrupting nodes and uses a deterministic function
for selecting which nodes to attack, based on the budget and public information about the PCN. 
Honest nodes are not aware of $\BB$ or the set of nodes $\XX$ attacked by the adversary, but use public information about the PCN to estimate the adversarial strategy and $\BB$ (e.g., as a fraction of the total PCN capacity). This in turn outputs a set of \textit{potentially} corrupted nodes, say $\Tilde{\XX}$, where likely $\XX \neq \Tilde{\XX}$.

\lukas{We choose a strategy to estimate $\Tilde{\XX}$, such that they are the optimal set of nodes for an adversary to corrupt, i.e., the largest fraction of payments is affected using a fixed budget, based on previous works \cite{tikhomirov2020quantitative,kapposEmpiricalAnalysisPrivacy2021,DBLP:journals/corr/abs-2103-08576,10.1145/3465481.3465761}. %
This approach allows the honest nodes to deny an adversary exactly this optimal placement within the graph by bypassing the nodes in $\Tilde{\XX}$. By denying adversaries the nodes where they can do the most damage, the overall security and privacy against on-path attacks is improved, while attacks become less profitable for the adversary. Moreover, if we establish a direct VC between two end-users, we effectively prevent on-path attacks \emph{regardless of the adversarial strategy}. 
There are many other strategies, our approach is modular, and can be used to study any other adversarial strategy to compute $\Tilde{\XX}$.} 
Our algorithms use $\Tilde{\XX}$ as part of their input, thus $\Tilde{\XX}$'s computation is modular in our design.

\paragraph{On-path attacks} For a path $s-u_1-\ldots-u_n-r$ the attacks on value privacy, on relationship anonymity, and the wormhole attack are defined as follows. 

\begin{asparaitem}
\item \textit{Value privacy}~\cite{malavolta2017concurrency}: PCN payments ensure that the transaction amount remains private to off-path corrupted users if there are only honest users along the path. This means, that if there are on-path corrupted users, value privacy does not hold anymore, as they can simply see the value and leak it to users not on the path. \textit{Preventing this attack:} For all segments $u_i-x_1-x_2-\ldots - x_\ell-u_j$ of the path from $s$ to $r$, where $u_i$, $u_j$ are not corrupted and $x_p$, $p=1,\ldots,\ell$ are corrupted, build a virtual channel from $u_i$ to $u_j$.

\item \textit{Relationship anonymity}~\cite{malavolta2017concurrency}: If an adversary controls two corrupted users $u_1$ and $u_n$, they can distinguish who is paying to whom. \textit{Preventing this attack:} If $u_i-x_1-x_2-\ldots - x_\ell-u_j$ is a segment of the path from $s$ to $r$, where $u_i,u_j$ are not corrupted, $x_p$, $p=1,\ldots,\ell$ are corrupted and $u_i \in \set{s, r} \lor u_j \in \set{s,r}$. If both $s$ and $r$ are part of such a segment, take one segment (there can be at most two) and build a virtual channel from $u_i$ to $u_j$. %

\item \textit{Wormhole attack}~\cite{malavoltaAnonymousMultiHopLocks2019}: In PCN payments, an adversary can prevent honest users from finalizing payments and effectively steal their fees. For this, the adversary needs to control corrupted nodes on both sides of one or more honest nodes along the path. \textit{Preventing this attack:} Identify all segments  $u_i-x_1-\ldots-x_\ell-y_1-\ldots-y_m-z_1-\ldots-z_n-u_j$ of the path from $s$ to $r$ where $u_i$, $u_j$ and $y_p$, $p = 1,\ldots,m$ are not corrupted and where $x_q$, $q = 1,\ldots,\ell$ and $z_r$, $r = 1, \ldots, n$ are corrupted. For each segment, build one of the following virtual channels: 
(i) between $u_i - y_1$
(ii) between $y_m - u_j$
(iii) between $x_\ell - z_1$.

\end{asparaitem}

\paragraph{Costs of VCs} Once opened, VCs can effectively reduce the fees of payments within a VPCN, as we explained in~\cref{sec:vc-background}. However, to create a VC, the endpoints need to pay an establishment fee $\feecreate$. Since VCs are currently not used, there is no fee model in practice which we can use. We therefore assume that $\feecreate$ of a VC over some path with capacity $\alpha$ to be the same as users would charge for forwarding a 
payment of amount $\alpha$ over that path. I.e., node $\userid_{j}$ charges $fee_{j} = f_{j,j+1} + p_{j,j+1} \cdot (\alpha + \sum_{k=j+1}^n fee_k)$, for $j=1,\ldots,n$. We discuss other potential fee models in~\cref{sec:discussion} and note that $\feecreate$ is modular in our model.

\paragraph{Optimization goal} 
Our objective is to set up virtual channels such that the cost for routing a set of transactions is minimized, and no transaction is traversing a path prone to an attack. 
Definition \ref{def:optgoal} consolidates our optimization goal and its hardness is proven in Theorem \ref{thm:NPhardness}.

\begin{definition}[\vpcn cost optimization]
\label{def:optgoal}
Given a \vpcn $\GG$, a set of transactions $\TT$, an estimated strategy of an on-path adversary to corrupt nodes and the estimated budget of the adversary $\BB$ for doing so, minimize the cost for routing the transactions in $\TT$, such that no transaction is traversing a path that is prone to a given attack.
If the estimation of the adversary's budget is $\BB=0$ our goal is  to minimize the routing fees.
\end{definition}

\begin{theorem}
\label{thm:NPhardness}
The \vpcn cost optimization problem is NP-hard.
\end{theorem}

\begin{proof}
We reduce an instance of the (NP-complete) minimum-length disjoint paths (MLDP) problem \cite{eilam1998disjoint} to an instance of our VPCN problem.
Consider an instance of the MLDP problem, i.e., an arbitrary directed graph $G = (V,E)$ such that $weight_G(u,v) = 1$, for all $(u,v) \in E$, and two source destination pairs $(s_1, d_1)$ and $(s_2, d_2)$ (two pairs are enough to render the problem NP-complete \cite{eilam1998disjoint}).

We now build an instance of the VPCN problem. We define $G' = (V, E\cup E')$, where $E' = \{(x,y) \,|\, (y,x) \in E \land (x,y) \notin E\}$, such that $weight_{G'}(x,y) = 1+\varepsilon$, $\varepsilon = 1/|E|$, if $(x,y)\in E$ and $weight_{G'}(x,y) = 0$ if $(x,y)\in E'$. 
Thus, for every pair of nodes $x,y$ in $G'$ either $\{(x,y), (y,x)\} \in G'$ or $x$ and $y$ are not adjacent.
The payment channels are hence defined as all the pairs of nodes $x,y$ in $G'$ such that $\{(x,y), (y,x)\} \in G'$, with capacity $weight_{G'}(x,y) + weight_{G'}(y,x)$, base fee equal to $\varepsilon$ and proportional fee equal to 0.
We consider the set of transactions, in the form of (source, destination, amount), to be $\{(s_1, d_1, 1), (s_2, d_2, 1)\}$.
We also, assume that creating virtual channels is not possible, as the problem only becomes harder by including them.
We set $c_{tr}$, the target percentage of successful transactions, to 1 (all should be executed).

A solution to the VPCN problem gives the minimum cost payment path for the two transactions.
This set of payment paths in $G'$ is using only edges that appear in $G$, as we set the capacity of the extra edges to zero and since the edge weights can accommodate for only one payment path, it does so with minimum length in $G$ and also the paths are edge disjoint (the capacity suffices for only one transaction).
Therefore a solution of the VPCN problem (payment paths) is exactly a solution of the minimum-length disjoint paths problem.
\end{proof}

\section{Exact algorithm}
\label{sec:ILPfull}

We present an exact solution to the VPCN cost optimization problem (Definition \ref{def:optgoal}) by modeling it as an Integer Linear Program (ILP). 
Our first challenge is to define the objective function to be optimized. We have three objectives: (a) minimize routing fees, 
(b) minimize virtual channel creation costs, and
(c) maximize successful transactions (either in number or in volume). 
We will define the objective function using items (a) and (b), and form the ILP as a minimization problem. Item (c) will be converted to a constraint (this is common in multi-objective optimization), requiring that the success ratio is above a threshold given in the input. Thus, different threshold values might yield different solutions.

The second challenge is to define the invariants that a solution should respect and, based on them, specify the ILP's variables and constraints. We identify the following invariants:
(i) at most one path is used for routing a transaction, 
(ii) the transaction success ratio should be above the given percentage,  
(iii) capacities of payment and virtual channels are respected,    
(iv) a VC between $i,j$ over $k$ should be bidirectional,     
(v) a VC is constructed if and only if it is used for routing a transaction or for constructing a higher-order VC,
(vi) payment paths prone to attacks of on-path adversaries are not selected in the ILP solution.
From a geometric point of view, the constraints define a set (polytope) of feasible solutions and an ILP solver outputs a feasible solution (if any) within this set that produces the minimum value for the objective function, i.e. the function that expresses objectives (a) and (b) as a linear combination of the variables. We will now define the ILP formally.

\paragraph{Input}
The input needed to define the ILP is a PCN (as defined in Section \ref{sec:background}) including all the payment channels and their attributes, a set of $T$ transactions $\TT = \{transaction_t = (s_t, d_t, trans_t)\}_{t \in [1,T]}$, i.e., (source, destination, amount), a constant $c_{tr} \in [0,1]$ indicating the required minimum success or volume ratio, i.e., if $c_{tr}=1$ all transactions must be executed, and the the set $\Tilde{\XX}$ of estimated (by the honest nodes) set of corrupted nodes.

Let $ch_{ij}.base\_fee$ ($f_i$ in Definition \ref{def:vpcn}) and $ch_{ij}.prop\_fee$ ($p_i$ in Definition \ref{def:vpcn}) denote the base and proportional forwarding fee of a payment ($ch_{ij} = pc_{ij}$) or a virtual channel ($ch_{ij} = vc_{ij}^k$, where $k$ is the intermediary node), and $pc_{ij}.capacity$ denotes the payment channel (PC) capacity ($\beta_i$ in Definition \ref{def:vpcn}). 
We set the virtual channel (VC) fees to be equal to those of the underlying initial payment channel, i.e., the fees of $vc_{ij}^k$ match those of $pc_{ik}$ if $vc_{ij}^k$ is built over $pc_{ik}$.

Since a virtual channel $vc_{ij}^k$ can be constructed over any combination of two adjacent payment or virtual channels $ch_{ik}$ and $ch_{kj}$, we assume a recursive structure of VCs and bound the levels of recursion by the input parameter $w$. Level-0 virtual channels are constructed over two adjacent payment channels. Level-$m$ virtual channels, $0 < m \leq w$, are constructed over a level-$(m-1)$ virtual channel and an adjacent payment or virtual channel of level at most $m-1$.
The VPCN that we provide as part of the ILP input will be a fusion of all PCs and all possible VCs, such that the ILP solver can decide which VCs to utilize.
To distinguish which VCs were used for payment paths or enforced for bypassing corrupted nodes, we formally define the input VPCN as a directed graph over the set of all nodes $V$ and two sets of edges (channels): $E_{PC}$ (PCs) and $E_{VC}$ (all possible VCs). 

We define each edge (channel) in $E_{VC}$ by the triple $(i,j,id)$, where $i,j$ are the endpoints and $id$ is a unique edge identifier. The edge id will allow us to distinguish two level-$m$ ($m$>0) VCs $vc_{ij}^k$ between $i$ and $j$ over $k$ built over different VCs, e.g. $vc_{kj}^u$ and $vc_{kj}^v$. We define $E_{PC} = \{(i,j,\langle i,j \rangle) \,|\, pc_{ij} \text{ exists}\}$. We then define $E_{VC}$ as the union of all possible VCs for each level 0 to $w$. We define $E_{VC}^0 = \{(i,j, ch_{ik}.id \circ ch_{kj}.id ) \,|\, ch_{ik} = (i,k, id) \land ch_{kj} = (k,j, id') \land ch_{ik},ch_{kj}\in E_{PC}\}$, where $\circ$ is a function that joins two ids to a unique new id, e.g. $id \circ id' = \langle id, id' \rangle$. For $1\leq m \leq w$, $E_{VC}^m = \{(i,j, ch_{ik}.id \circ ch_{kj}.id ) \,|\, ch_{ik} = (i,k, id) \land ch_{kj} = (k,j, id') \land ch_{ik},ch_{kj}\in E_{PC}\cup (\cup_{\ell=0}^{m-1} E_{VC}^\ell) \land \{ch_{ik}, ch_{kj}\}\cap E_{VC}^{m-1}\neq \emptyset\}$.
For example, the edge id can be a breakdown of all edges (channels) building it: the id of a PC between $i,j$ is $\langle i,j \rangle$, the id of a level-0 VC between $i,j$ over $k$ is $\langle\langle i,k \rangle, \langle k,j \rangle\rangle$, and the id of a level-$m$ VC consisting of two channels $ch_x$ and $ch_y$ is $\langle ch_x.id, ch_y.id\rangle$, where $m\leq w$.

The recursion depth $w$ can make $E_{VC}$ and hence the ILP size exponential. For example, for $w=0$, we need to consider $\bigO({n \choose 2}) = \bigO(n^2)$ paths of size 2 for constructing all possible level-0 VCs. However, when considering $w=\Theta(n)$, the size of $E_{VC}$ becomes exponential, as it is proportional to $\bigO(\sum_{k=2}^n {n \choose k}) = \bigO(2^n)$.

\paragraph{Constants, variables, and macros}
We will use three sets of integer variables. The first set includes binary variables that indicate that $transaction_t$ is routed via path $P$ ($path_P(trans_t)$), the second set indicates the capacity of a virtual channel ($vc_{ij}^k.capacity$), and the third set indicates whether a virtual channel exists ($exists\_vc_{ij}^k$).

Let $\PP(s_t, d_t)$ be a list of all the paths from a sender $s_t$ to a receiver $d_t$ for $transaction_t$ in the graph $(V, E_{PC}\cup E_{VC})$.
The variable $path_P(trans_t)\in\{0,1\}$ indicates whether $transaction_t$ is routed through path $P\in \PP(s_t, d_t)$.
This set of variables is exponential on the number of nodes, but our goal here is to design an exact solution to an NP-complete problem, thus this is expected. The exact solution is necessary step before designing fast exact solution implementations or approximations (e.g. ILP relaxations and rounding rules).
For a channel $ch_{ij}$ and $transaction_t$, we define the macro $used(ch_{ij},t) = \sum_{P\in \PP(s_t, d_t)} path_P(trans_t)\cdot In(ch_{ij}, P)$, where $In(ch_{ij}, P)$ is a constant indicating whether $ch_{ij}\in P$. When $ch_{ij}$ is used by a payment route for $transaction_t$, then $used(ch_{ij},t)$ is 1 (only one path is used for $transaction_t$ due to constraint C1), and otherwise, it is 0.
Let 
$routing\_fee(t, P,ch_{ij})$, $ch_{ij}\in \{pc_{ij}, vc_{ij}^k\}$
be the routing fees charged to channel $ch_{ij}\in P$ for $transaction_t$. We note that $routing\_fee(t$, $P,ch_{ij})$ is computed as in Section~\ref{sec:pcns-background} if $ch_{ij}\in P$ and is zero otherwise.
Let $routing\_cost_{ch_{ij}} =
\sum_{t=1}^{T}\sum_{P\in \PP(s_t,d_t)} routing\_fee(t,P,ch_{ij}) \cdot path_P(trans_t)$  
be the routing fees that are charged for the transactions that traverse channel $ch_{ij} \in \{pc_{ij}, vc_{ij}^k\}$.

Paths including nodes in $\Tilde{\XX}$, i.e., estimated to be corrupted, should not be used for routing payments, thus we exclude those paths from $\cup_{t=1}^T\PP(s_t, d_t)$.
For instance, for value privacy, we exclude every path $P$ such that $x\in P$, for all $x \in \Tilde{\XX}$.
Thus, all remaining input paths can be safely selected by the ILP solver.

We denote the capacity of a virtual channel $vc_{ij}^k \in E_{VC}$ with $vc_{ij}^k.capacity$.
We define the virtual channel creation cost as  
$vc_{ij}^k\_creation\_cost = exists\_vc_{ij}^k \cdot vc_{ij}^k.base\_fee +  vc_{ij}^k.prop\_fee \cdot vc_{ij}^k.capacity$, where $exists\_vc_{ij}^k$ is a binary variable indicating if $vc_{ij}^k$ exists.
If $vc_{ij}^k$ is used for routing transactions ($exists\_vc_{ij}^k=1 \land vc_{ij}^k.capacity>0$), then the creation cost is $vc_{ij}^k.base\_fee +  vc_{ij}^k.prop\_fee \cdot vc_{ij}^k.capacity$.
Due to constraint C5, if $vc_{ij}^k$ is not used in a payment path, $exists\_vc_{ij}^k = 0$ and because $vc_{ij}^k.creation\_cost$ appears in the objective function, which we want to minimize, $vc_{ij}^k.capacity$ will be reduced to 0 in any minimal solution. Thus $vc_{ij}^k.creation\_cost = 0$ when $vc_{ij}^k$ is not used in a payment path.

\paragraph{Objective}
The objective is to minimize routing and virtual channel creation costs:
		$\min \sum_{pc_{ij}\in E_{PC}} routing\_cost_{pc_{ij}} + \sum_{vc_{ij}^k \in E_{VC}} (routing\_cost_{vc_{ij}^k} +  vc_{ij}^k\_creation\_cost)$.

\setcounter{concounter}{0}
\paragraph{Constraints} We define five constraints that collectively express the invariants:\\ 
\newcon At most one path can be used for routing a transaction:\\  $\sum_{P \in \PP(s_t, d_t)} path_P(trans_t) \leq 1$, $\forall t\in [1,T]$.\\
\newcon The percentage of successful transactions or volume should be at least $c_{tr} \in [0,1]$. To define this constraint we first define the sum $\sum_{P\in \PP(s_t,d_t)} path_P(trans_t)$ that is 1 when a transaction is successful (only one path routes the transaction) and 0 otherwise (no path is selected). Note that this sum is binary due to C1. We then express the constraint as follows: 
$\sum_{t=1}^T trans_t \cdot \sum_{P\in \PP(s_t,d_t)} path_P(trans_t) \geq c_{tr}\sum_{t=1}^T trans_t$, where $c_{tr}$ is the success volume ratio. In case $c_{tr}$ is the minimum percentage of successful transactions, this constraint becomes $\sum_{t=1}^T \sum_{P\in \PP(s_t,d_t)} path_P(trans_t) \geq c_{tr}T$.\\
\newcon Virtual and payment channel capacities should be respected. The load on each channel is the sum of routing costs, transaction amounts, as well as the VC creation costs and capacities for higher order VCs in $E_{VC}$ that use the channel (we denote those with $vc_{ix}^j = (ch_{ij}, \bullet)$, where $\bullet$ is any channel between $j$ and $x$ in $E$ that can form $vc_{ix}^j$). Thus for every $ch_{ij} \in E = E_{PC}\cup E_{VC}$ we require that
\begin{center}
$routing\_cost_{ch_{ij}} + trans\_amount(ch_{ij}) +$\\ $\sum_{vc_{ix}^j \in E_{VC}: vc_{ix}^j = (ch_{ij}, \bullet)} (vc_{ix}^j.creation\_cost + vc_{ix}^j.capacity) \leq ch_{ij}.capacity$
\end{center}
where $trans\_amount(ch_{ij}) = \sum_{t=1}^T trans_t \cdot used(ch_{ij}, t)$.
If possible by the capacities, C2 forces the ILP solver to set some $path_P(trans_t)$ variables to 1, i.e., some paths are selected for routing transactions and thus the 0 solution (no successful transaction) is prohibited for the inputs where a feasible solution exists. 
Moreover, if a VC is in a path selected by the ILP solution for routing a payment, then the routing cost and the transaction amount that are charged to this VC's capacity are positive, and thus the VC capacity is positive and lower bounded by this amount. Since VC capacities are part of the objective function, any minimal solution will assign the minimal VC capacity for routing transactions or creating higher order VCs. Similarly, VC capacity will be 0 for VCs that are not used.\\
\newcon A VC between nodes $i,j$ over $k$ must exist in both directions $(i,k,j)$ and $(j,k,i)$:\\
$exists\_vc_{ij}^k = exists\_vc_{ji}^k$\\
\newcon A VC exists, only if it is used for routing a transaction or to construct a VC of higher recursive order:\\
$exists\_vc_{ij}^k \leq \sum_{t=1}^T used(vc_{ij}^k,t) + \sum_{vc_{sr}^\ell\in rec(vc_{ij}^k)} exists\_vc_{sr}^\ell$
and\\
$exists\_vc_{ij}^k \geq \frac{1}{T + |E_{VC}|}\left(\sum_{t=1}^T used(vc_{ij}^k,t)+\right.$\\
$~~~\qquad\qquad\qquad\qquad\qquad\quad\left. \sum_{vc_{sr}^\ell\in rec(vc_{ij}^k)} exists\_vc_{sr}^\ell\right)$,\\ where $rec(vc_{ij}^k)$ includes all $vc_{sr}^\ell \in E_{VC}$ that are built over $vc_{ij}^k$. 
The first inequality enforces $exists\_vc_{ij}^k = 0$ if the VC is not used and the second one enforces $exists\_vc_{ij}^k = 1$ if the VC is used. 
Note that, $exists\_vc_{ij}^k$ appears in $vc_{ij}^k.creation\_cost$ as a factor of $vc_{ij}^k.base\_fee$, making the VC creation cost calculation accurate.
We give an example run of the ILP in Appendix \ref{sec:ILPexample}.

\paragraph{Output}
We can determine all created VCs by checking $exists\_vc_{ij}^k$ and if $transaction_t$ is successful by the value of $\sum_{P\in \PP(s_t, d_t)} path_P(trans_t) \in \{0,1\}$.  

\paragraph{Computational complexity} 
The ILP has exponentially many variables and constraints (asymptotically) since it needs to decide which subset of all paths minimizes the objective.
We implemented the ILP using Python and Gurobi \cite{gurobi}. We were able to run it with at most 15 nodes, 30 channels, and 5 transactions (cf. Appendix \ref{sec:ILPexperiments} and Figure \ref{fig:ILPexecTime}).
One way to make the ILP solution computation more tractable is to restrict $w=\bigO(1)$ (VC recursion bound) and to limit the number of possible payment paths per transaction to be polynomially many.

\subsection{ILP Example} 
\label{sec:ILPexample}

We apply the ILP to the example graph of Figure~\ref{fig:graph_example} with the same input. In this example we also assume that the capacities are enough for routing all transactions and opening the VCs shown in the figure ($VC_1$, $VC_2$, $VC_3$), it is possible to build level-0 and level-1 VCs, the minimum success volume ratio is 1, and that $\Tilde{\XX} = \{H_1\}$. 
We first remove from the ILP input all paths containing $H_1$.
The ILP will output $VC_1$, $VC_2$, $VC_3$ as the paths used for A-to-B, B-to-C, and A-to-C transactions, respectively. These payment paths are the cheapest ones for the corresponding transactions and it is possible to build them given the underlying PC capacities. Note that any other path would be longer and more costly due to linear routing fees.

We now check how all constraints are respected. According to C1, only the payment paths $VC_1$, $VC_2$, $VC_3$ will be used ($path_P(trans_t)=1$ only for those paths and 0 for all others). C2 will be satisfied since all transactions are successful. C3 is satisfied because we assumed there are enough capacities to build all three VCs and route all transactions. In fact, since the VC capacities are minimized in the objective function, they will be just enough to route the input transactions. C4 will force $VC_1$, $VC_2$, $VC_3$ exist in both directions. The first inequality of C5 will force all VCs in $E_{VC}\setminus \{VC_1, VC_2, VC_3\}$ to not exist since the right side of the inequality will be 0, while the second inequality will force $VC_1$, $VC_2$, $VC_3$ to exist since the right side of the inequality will be a positive value in $(0,1]$.

\subsection{ILP experiments}
\label{sec:ILPexperiments}

\begin{figure}[htbp]
  \centering
  \includegraphics[width=0.49\textwidth]{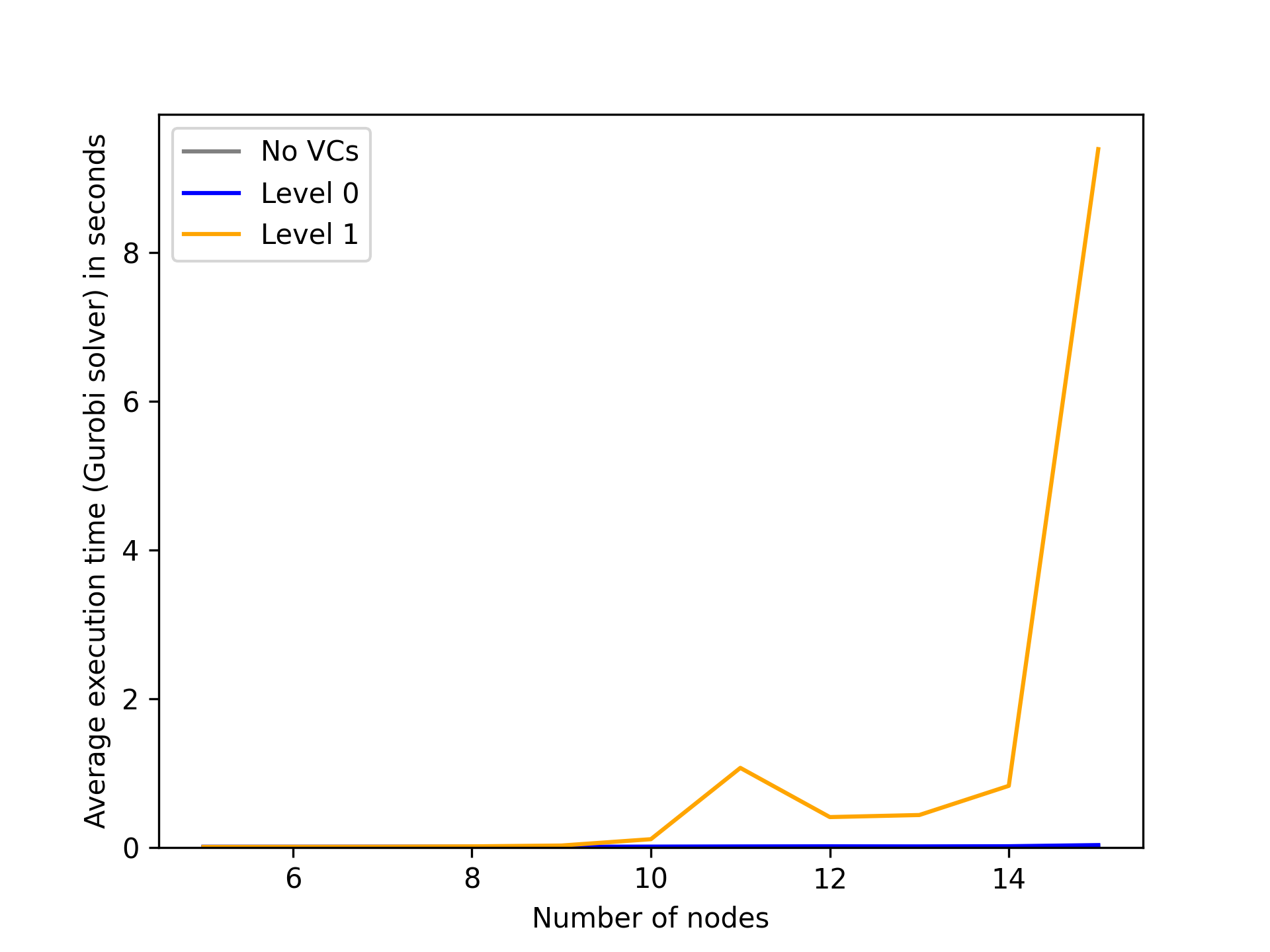}
  \hfill
  \includegraphics[width=0.49\textwidth]{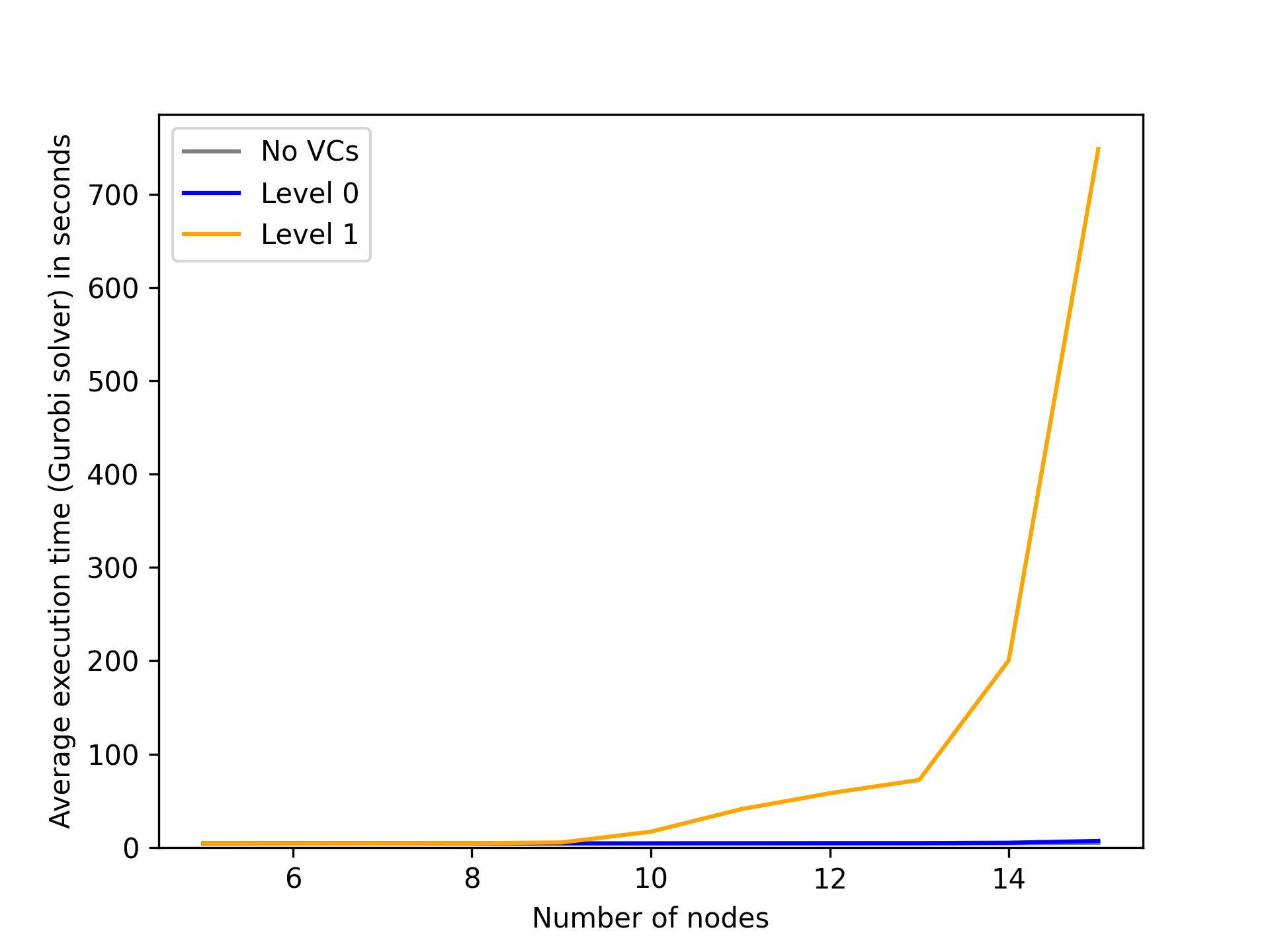}
  \caption{Average execution time to number of nodes, with 5 transactions. On the left we show the execution time for the Gurobi Solver and on the right the execution time for building the ILP model. Both show exponential growth and the solver started to crash for some graphs with 15 nodes.}
  \label{fig:ILPexecTime}
\end{figure}
We implemented the ILP with Gurobi \cite{gurobi} and experimented with the size of the graphs it can solve. The github repository of our implementation and experimental evaluation can be found in \cite{yannikrepo}. We experimented both with random graphs and with a custom heuristic that produces random graphs that resemble the Lightning Network, i.e. graphs with dense core and sparse boundary. To compute the latter, we extracted the probability distribution of the percentage of nodes to which a node is connected to, i.e. $d_i/N$, where $d_i$ is the degree of node $i$ and $N$ is the number of nodes in the Lightning Network snapshot we used. Then, given the size of the new (smaller) graph, say $k$, we sampled $k$ values uniformly at random from that distribution. We fixed the sample graph edges to nodes ratio to 2:1, and when needed we added some more edges randomly from the leaf nodes to maintain the desired characteristics of the LN. Both for random graphs and for the graphs computed with our heuristic, we computed each case by running each experiments 25 times and taking the average value. We show our findings in Figure \ref{fig:ILPexecTime}.

\section{An efficient greedy algorithm}
\label{sec:algsandanalysis}

\label{sec:greedy}

In the following, we present an efficient greedy algorithm 
with low running time while ensuring high-quality channel allocations
(see also the upcoming evaluation in \cref{sec:heuristics}).
Before investigating our overarching optimization goal of \cref{sec:model}, where we aim to prevent any attacks by on-path adversaries and minimize routing fees, our algorithm will optimize for the following goals individually: preventing
\begin{enumerate*}[label=(\roman*)]
    \item \label{enumitem:optgoal1} relationship anonymity attacks,
    \item \label{enumitem:optgoal2} wormhole attacks,
    \item \label{enumitem:optgoal3} value privacy attacks, and
    \item \label{enumitem:optgoal4} minimizing routing fees
\end{enumerate*}.

Our greedy algorithm is given as input the PCN, the payments that are to be carried out, and the optimization goal, which can be one of the optimization goals \labelcref{enumitem:optgoal1,enumitem:optgoal2,enumitem:optgoal3,enumitem:optgoal4}. Each payment consists of a \emph{sender}, a \emph{receiver}, some \emph{value} and the number of times this payment is carried out (\emph{repetition}). The algorithm will iterate over the list of payments and for each payment, try to find the cheapest path(s) in terms of fees using Dijkstra's algorithm with enough capacity to route it (abstracted as \emph{generate\_paths}). Then, the algorithm will compute which nodes to bypass (abstracted as \emph{compute\_nodes\_to\_bypass}) in order to prevent on-path adversary attacks (optimization goals \labelcref{enumitem:optgoal1,enumitem:optgoal2,enumitem:optgoal3}) or to minimize routing fees (optimization goal~\labelcref{enumitem:optgoal4}). Finally, the algorithm constructs virtual channels to bypass these nodes and conducts the payments (abstracted as \emph{construct\_vcs} and \emph{conduct\_payments}). \lukas{This algorithm can be applied locally, by individual nodes who do not know about any payments other than their own.}
We give a high-level pseudocode of this approach in \cref{alg:greedymaindbody}.

\algrenewcommand\algorithmicindent{0.7em}%
\begin{algorithm}
  \caption{High level greedy algorithm}
    \label{alg:greedymaindbody}
  \begin{algorithmic}[1]
  \footnotesize
    \Statex
    \Function{greedy\_vc\_algorithm}{pcn, payments, optimization\_goal}
        \For{(sender, receiver, value, repetition) in payments}
            \State{\pseudocomment{find cheapest path(s) with capacity to route each repeat payment}}%
            \Let{paths}{generate\_paths(pcn, sender, receiver, value, repetition)}
            \For{(path, amount) in paths}       \State{\pseudocomment{Compute the nodes which to bypass, based on optimization\_goal}}
                \Let{nodes\_to\_bypass}{compute\_nodes\_to\_bypass(path, optimization\_goal)}
                \State{\pseudocomment{Build virtual channels over these nodes}}
                \Let{(pcn, new\_path)}{construct\_vcs(pcn, path, nodes\_to\_bypass, amount)}
                \State{\pseudocomment{Conduct payments and update channel balances}}
                \Let{pcn}{conduct\_payments(pcn, new\_path, amount)}
            \EndFor
      \EndFor
    \EndFunction
  \end{algorithmic}
\end{algorithm}

While we will present concrete pseudocode for implementing \emph{compute\_nodes\_to\_bypass} to achieve each optimization goal 
\labelcref{enumitem:optgoal1,enumitem:optgoal2,enumitem:optgoal3,enumitem:optgoal4} 
\iftechreport
in \cref{alg:bypassVP,alg:bypassRA,alg:bypassWH,alg:bypassFees} in \cref{sec:algorithms}, 
\else
in Algorithms 5 to 8 in \cite[Section V-C]{threebirds-IFIP-TR}.
\fi
we give a short outline here. 
For relationship anonymity, it is sufficient to greedily bypass 
corrupted nodes adjacent to either the sender or the receiver, along short paths. To prevent the wormhole attack, corrupted nodes need to be bypassed such that there are no honest nodes encased by corrupted nodes. For value privacy, all corrupted nodes need to be bypassed.

Regarding fee optimization, we recall that the fee for opening a VC of capacity $\alpha$ is the same as routing a payment of amount $\alpha$ via that path. In our model, this means that it is cheaper to create a VC between sender and receiver, as soon as we carry out a payment on a path more than once. %
In 
\cref{sec:strategy_relation},
we show that there is a synergy between these goals and that opening VCs is beneficial to all goals. 
\iftechreport
Our greedy algorithm runs efficiently on commodity hardware, as we show now.
\else
Our greedy algorithm runs efficiently on commodity hardware. 
The time complexity of \cref{alg:greedymaindbody} is $\Theta(T \cdot (|E| + |V|\log|V| + D))$, 
where $D$ is the diameter 
(cf. \cite[Section V-A]{threebirds-IFIP-TR}).
\fi

\subsection{Runtime analysis of greedy algorithm}
\label{sec:greedy-runtime}
Our greedy algorithm is efficient enough, such that we can run the experiments we conduct in \cref{sec:heuristics} on a Lightning Network snapshot on commodity hardware. Dijkstra's algorithm has a runtime of $\Theta(|E| + |V|\log|V|)$. We need to run Dijkstra's algorithm to find the shortest path for each payment because the PCN topology and the channel capacity change after each payment and VC construction, and the payment amount is different. Additionally, for each payment, we need to traverse each edge to (temporarily) remove edges that do not have enough capacity to route each payment. Finally, identifying which nodes to bypass, creating the VC, and routing the payment are all linear in the number of nodes on the payment path. Thus, if $T$ is the number of all payments and $D$ the network diameter when considering only PCs (and a bound to the maximum length of a payment path), the total complexity of \cref{alg:greedymaindbody} is $\Theta(T \cdot (|E| + |V|\log|V| + D))$.

\subsection{Example}
\label{sec:examplesolution}

To demonstrate our approach, we find the solutions of both the ILP, i.e., the optimal solution computed by an integer linear program 
\iftechreport
(cf. Section~\ref{sec:ILPfull}), 
\else
(cf. Section~\ref{sec:ILPsummary}), 
\fi
and the greedy algorithm on a small example graph. 
The graph is shown in \cref{fig:graph_example} and consists of two hub-like nodes $H_1$ and $H_2$, and four client nodes $A$, $B$, $C$ and $D$. We say that each channel charges a base fee of $1$, a proportional fee of $0.001$, and has a capacity of $10$k, distributed evenly among both users. The transactions that are executed have all values of $10$; there are three transactions from $A$ to $C$, one transaction from $A$ to $B$, and one transaction from $B$ to $C$. Further, we assume that $H_1$ is a malicious node. 
We chose this graph because it resembles the hub-and-spoke topology of the Lightning Network~\cite{zabka2022centrality}. The transaction values and fees are chosen mainly for readability: In the total fees below, the integer part of the number represents the sum of the base fees, and the fractional part is the sum of the proportional fees.
\begin{figure}
    \centering
    \resizebox{0.6\columnwidth}{!}{
        \tikzset{every picture/.style={line width=0.75pt}} %

\begin{tikzpicture}[x=0.75pt,y=0.75pt,yscale=-1,xscale=1]

\draw   (140,145) .. controls (140,136.72) and (146.72,130) .. (155,130) .. controls (163.28,130) and (170,136.72) .. (170,145) .. controls (170,153.28) and (163.28,160) .. (155,160) .. controls (146.72,160) and (140,153.28) .. (140,145) -- cycle ;
\draw   (210,105) .. controls (210,96.72) and (216.72,90) .. (225,90) .. controls (233.28,90) and (240,96.72) .. (240,105) .. controls (240,113.28) and (233.28,120) .. (225,120) .. controls (216.72,120) and (210,113.28) .. (210,105) -- cycle ;
\draw   (360,145) .. controls (360,136.72) and (366.72,130) .. (375,130) .. controls (383.28,130) and (390,136.72) .. (390,145) .. controls (390,153.28) and (383.28,160) .. (375,160) .. controls (366.72,160) and (360,153.28) .. (360,145) -- cycle ;
\draw   (280,145) .. controls (280,136.72) and (286.72,130) .. (295,130) .. controls (303.28,130) and (310,136.72) .. (310,145) .. controls (310,153.28) and (303.28,160) .. (295,160) .. controls (286.72,160) and (280,153.28) .. (280,145) -- cycle ;
\draw   (210,185) .. controls (210,176.72) and (216.72,170) .. (225,170) .. controls (233.28,170) and (240,176.72) .. (240,185) .. controls (240,193.28) and (233.28,200) .. (225,200) .. controls (216.72,200) and (210,193.28) .. (210,185) -- cycle ;
\draw   (60,145) .. controls (60,136.72) and (66.72,130) .. (75,130) .. controls (83.28,130) and (90,136.72) .. (90,145) .. controls (90,153.28) and (83.28,160) .. (75,160) .. controls (66.72,160) and (60,153.28) .. (60,145) -- cycle ;
\draw    (90,145) -- (140,145) ;
\draw    (310,145) -- (360,145) ;
\draw    (240,180) -- (280,150) ;
\draw    (210,180) -- (170,150) ;
\draw    (210,110) -- (170,140) ;
\draw    (280,140) -- (240,110) ;
\draw [color={rgb, 255:red, 39; green, 0; blue, 255 }  ,draw opacity=1 ]   (75,130) .. controls (155,46) and (294,46) .. (375,130) ;
\draw [color={rgb, 255:red, 126; green, 211; blue, 33 }  ,draw opacity=1 ]   (90,140) .. controls (133,104) and (146,100) .. (210,100) ;
\draw [color={rgb, 255:red, 126; green, 211; blue, 33 }  ,draw opacity=1 ]   (240,100) .. controls (306,102) and (319,104) .. (360,140) ;

\draw (146,139) node [anchor=north west][inner sep=0.75pt]  [color={rgb, 255:red, 255; green, 0; blue, 0 }  ,opacity=1 ] [align=left] {$\displaystyle H_{1}$};
\draw (218,98) node [anchor=north west][inner sep=0.75pt]   [align=left] {$\displaystyle B$};
\draw (369,139) node [anchor=north west][inner sep=0.75pt]   [align=left] {$\displaystyle C$};
\draw (286,139) node [anchor=north west][inner sep=0.75pt]   [align=left] {$\displaystyle H_{2}$};
\draw (217,178) node [anchor=north west][inner sep=0.75pt]   [align=left] {$\displaystyle D$};
\draw (67,139) node [anchor=north west][inner sep=0.75pt]   [align=left] {$\displaystyle A$};
\draw (209,47) node [anchor=north west][inner sep=0.75pt]  [color={rgb, 255:red, 39; green, 0; blue, 255 }  ,opacity=1 ] [align=left] {VC$\displaystyle _{3}$};
\draw (141,108) node [anchor=north west][inner sep=0.75pt]  [color={rgb, 255:red, 126; green, 211; blue, 33 }  ,opacity=1 ] [align=left] {VC$\displaystyle _{1}$};
\draw (279,108) node [anchor=north west][inner sep=0.75pt]  [color={rgb, 255:red, 126; green, 211; blue, 33 }  ,opacity=1 ] [align=left] {VC$\displaystyle _{2}$};

\end{tikzpicture}
    }
    \caption{Results of the ILP and greedy algorithms run on a small sample graph. 
    }
    \label{fig:graph_example}
\end{figure}

In \cref{fig:graph_example}, we show the VCs constructed 
by the greedy approach in color and note that they are the same as in the ILP  (optimal solution).
The optimal solution is to build a virtual channel $\text{VC}_1$ of capacity $40$ between $A$ and $B$, another virtual channel $\text{VC}_2$ of capacity $40$ between $B$ and $C$, and finally a third virtual channel $\text{VC}_3$ of capacity $30$ between $A$ and $C$. Finally, these VCs are used for routing the transactions. The total cost on fees is $3.11$, which comes from three times the base fee of $1$ (which is $3$) and three times the proportional fee (twice for capacity $40$, once for capacity $30$), which is around $.11$ for creating the VCs. Then, all sender-receiver pairs are directly connected, so there are no additional routing fees.

In the greedy approach, the same VCs are constructed. Since this algorithm greedily creates the best VCs for each sender-receiver pair individually, $\text{VC}_1$ does not have enough capacity and has to be constructed twice. This incurs an extra base fee of value $1$. Since the overall VC capacity does not change, the proportional fee of $.11$ remains, totaling $4.11$ in fees. 

Finally, if we look at the fees of routing these payments without any s, the total amount spent on fees is $11.11$. Since we now need to route the payment from $A$ to $C$ three times over the path, the intermediaries charge a base fee each time, resulting in $9$ coins alone. Furthermore, the payments without VCs are prone to value privacy attacks by $H_1$. %

\subsection{Algorithms}
\label{sec:algorithms}

In this section, we give pseudocode definitions for the algorithms used to conduct our empirical evaluation. \cref{alg:heurisitic1} describes the algorithm for computing the cost ratio for preventing attacks. \cref{alg:heurisitic2} describes the algorithm for computing the cost ratio for the goal of optimizing fees. \cref{alg:heurisitic3} is the subprocedure to compute the corrupted nodes.

Further, we present the algorithms that return the sets of nodes that need to be bypassed to prevent value privacy attacks (\cref{alg:bypassVP}), relationship anonymity attacks (\cref{alg:bypassRA}), wormhole attacks (\cref{alg:bypassWH}) and to optimize fees (\cref{alg:bypassFees}).

\begin{algorithm}
  \caption{Compute cost ratio of bypassing corrupted nodes}
    \label{alg:heurisitic1}
  \begin{algorithmic}[1]
    \Statex
    \Function{compute\_cost\_ratio}{pcn, adv\_budget, val\_range, n, repetition}
      \Let{pcn\_vc}{copy(pcn)} \pseudocomment{Copy the PCN graph for the VC simulation}
      \Let{c}{compute\_corrupted\_nodes(pcn, adv\_budget)} \pseudocomment{compute the most profitable corrupted nodes to a given adversary budget}
      \Let{payments}{generate\_payments(pcn, val\_range, n, repetition)} \pseudocomment{generates n random payments with a random value in val\_range in a PCN graph}
      
      \Let{establish\_cost\_vc}{0}
      \Let{route\_cost\_vc}{0}
      \Let{unsuccessful\_vc}{0}
      \Let{route\_cost\_pcn}{0}
      \Let{unsuccessful\_pcn}{0}
      
      \For{(path, value, repetition) in payments}
        \Let{cost, new\_path}{bypass\_corrupted\_nodes(c, path, value, repetition, pcn\_vc)}
        \Let{establish\_cost\_vc}{establish\_cost\_vc + cost}
        \For{$i\gets 0; i<$ repetition $; i++$}
            \Let{success, fees\_vc}{conduct\_payment(pcn\_vc, new\_path, value)} \pseudocomment{executes the payment over the given path}
            \If{success}
              \Let{route\_cost\_vc}{route\_cost\_vc + fees\_vc}
            \Else
                \Let{unsuccessful\_pcn}{unsuccessful\_pcn + 1}
            \EndIf
            \Let{success, fees\_pcn}{conduct\_payment(pcn, path, value)} \pseudocomment{executes the payment over the given path}
            \If{success}
              \Let{route\_cost\_pcn}{route\_cost\_pcn + fees\_pcn}
            \Else
              \Let{unsuccessful\_pcn}{unsuccessful\_pcn + 1}
            \EndIf
        \EndFor
      \EndFor
      \State \Return{$\frac{\text{establish\_cost\_vc + route\_cost\_vc}}{\text{route\_cost\_pcn}}$}
    \EndFunction
  \end{algorithmic}
\end{algorithm}

\begin{algorithm}
  \caption{Compute corrupted nodes}
    \label{alg:heurisitic3}
  \begin{algorithmic}[1]
    \Statex
    \Function{compute\_corrupted\_nodes}{pcn, adv\_budget}
      
      \Let{payments}{generate\_payments(pcn, val\_range, n, repetition)} \pseudocomment{generates n random payments}
      
      \Let{node\_map<node,int>}{\{\}}
      \For{(path, value, repetition) in payments}
        \For{node in path} 
            \Let{node\_map[node]}{node\_map[node]+1}
        \EndFor
    \EndFor
            
      \Let{cost\_benefit\_map<node,int>}{\{\}}
      \For{node in node\_map.keys()}
        \Let{cost\_benefit\_map[node]}{$\frac{\text{(node\_map[node]/n)}}{\text{(cost(node)/adv\_budget)}}$}
    \EndFor\Let{corrupted\_nodes}{[]}
      \For{(node,cost) in cost\_benefit\_map.sort\_by\_value(desc)}
        \If{adv\_budget $\geq$ cost(node)}
            \Let{corrupted\_nodes}{corrupted\_nodes.append(node)}
            \Let{adv\_budget}{adv\_budget - cost(node)}
        \EndIf
      \EndFor

      \State \Return{corrupted\_nodes}
    \EndFunction
  \end{algorithmic}
\end{algorithm}

\begin{algorithm}
  \caption{Compute cost ratio for saving fees}
    \label{alg:heurisitic2}
  \begin{algorithmic}[1]
    \Statex
    \Function{compute\_cost\_ratio}{pcn, val\_range, n, repetition}
      \Let{pcn\_vc}{copy(pcn)} \pseudocomment{Copy the PCN graph for the VC simulation}
      \Let{payments}{generate\_payments(pcn, val\_range, n, repetition)} \pseudocomment{generates n random payments with a random value in val\_range in a PCN graph}
      
      \Let{establish\_cost\_vc}{0}
      \Let{route\_cost\_vc}{0}
      \Let{unsuccessful\_vc}{0}
      \Let{route\_cost\_pcn}{0}
      \Let{unsuccessful\_pcn}{0}
      
      \For{(path, value, repetition) in payments}
        \Let{cost, new\_path}{open\_profitable\_vcs(c, path, value, repetition, pcn\_vc)}
        \Let{establish\_cost\_vc}{establish\_cost\_vc + cost}
        \For{$i\gets 0; i<$ repetition $; i++$}
            \Let{success, fees\_vc}{conduct\_payment(pcn\_vc, new\_path, value)} \pseudocomment{executes the payment over the given path}
            \If{success}
              \Let{route\_cost\_vc}{route\_cost\_vc + fees\_vc}
            \Else
                \Let{unsuccessful\_pcn}{unsuccessful\_pcn + 1}
            \EndIf
            \Let{success, fees\_pcn}{conduct\_payment(pcn, path, value)} \pseudocomment{executes the payment over the given path}
            \If{success}
              \Let{route\_cost\_pcn}{route\_cost\_pcn + fees\_pcn}
            \Else
              \Let{unsuccessful\_pcn}{unsuccessful\_pcn + 1}
            \EndIf
        \EndFor
      \EndFor
      \State \Return{$\frac{\text{establish\_cost\_vc + route\_cost\_vc}}{\text{route\_cost\_pcn}}$}
    \EndFunction
  \end{algorithmic}
\end{algorithm}

\begin{algorithm}
  \caption{Generating set of nodes to bypass for preventing VP}
    \label{alg:bypassVP}
  \begin{algorithmic}[1]
    \Statex
    \Function{compute\_bypass\_nodes\_VP}{path, corrupted\_nodes}
      
      \Let{nodes\_vp\_path}{\{\}}
      \For{node in path}
        \If{node in corrupted\_nodes}
            \Let{nodes\_vp\_path}{nodes\_vp\_path $\cup$ \{node\}}
        \EndIf
    \EndFor
            
      \State \Return{nodes\_vp\_path}
    \EndFunction
  \end{algorithmic}
\end{algorithm}

\begin{algorithm}
  \caption{Generating set of nodes to bypass for preventing RA}
    \label{alg:bypassRA}
  \begin{algorithmic}[1]
    \Statex
    \Function{compute\_bypass\_nodes\_RA}{path, corrupted\_nodes}
      \State Let nodes\_ra\_path\_l be the set of connected nodes in corrupted\_nodes adjacent to the sender path[0]
      \State Let nodes\_ra\_path\_r be the set of connected nodes in corrupted\_nodes adjacent to the receiver path[length(path)-1]

        \If{lenght(nodes\_ra\_path\_l) < length(nodes\_ra\_path\_r)}
            \State \Return{nodes\_ra\_path\_l}
        \EndIf
            
      \State \Return{nodes\_ra\_path\_r}
    \EndFunction
  \end{algorithmic}
\end{algorithm}

\begin{algorithm}
  \caption{Generating set of nodes to bypass for preventing WH}
    \label{alg:bypassWH}
  \begin{algorithmic}[1]
    \Statex
    \Function{compute\_bypass\_nodes\_WH}{path, corrupted\_nodes}
    \Let{nodes\_wh\_path}{\{\}}
    \For{while there exist honest nodes in path that is surrounded by corrupted nodes}
      \State Let hon\_nodes be one of these honest nodes surrounded by corrupted nodes
      \State Let wh\_l be the set of connected nodes in corrupted\_nodes adjacent to the left of hon\_nodes
      \State Let wh\_r be the set of connected nodes in corrupted\_nodes adjacent to the right of hon\_nodes

        \If{lenght(wh\_l) < length(wh\_r)}
            \State Remove wh\_l from path
            \State Add wh\_l to nodes\_wh\_path
        \Else
            \State Remove wh\_r from path
            \State Add wh\_r to nodes\_wh\_path
        \EndIf
        \Comment{Note that one could also bypass hon\_nodes, but we do not for simplicty here}
    \EndFor
    
      \State \Return{nodes\_wh\_path}
      
    \EndFunction
  \end{algorithmic}
\end{algorithm}

\begin{algorithm}
  \caption{Generating set of nodes to bypass for optimizing fees}
    \label{alg:bypassFees}
  \begin{algorithmic}[1]
    \Statex
    \Function{optimize\_fees}{path, corrupted\_nodes}
      
      \Let{nodes\_fees\_path}{path}
      \State Remove first and last element from nodes\_fees\_path

      \State \Return{nodes\_fees\_path}
    \EndFunction
  \end{algorithmic}
\end{algorithm}

\subsection{Synergy among objectives}
\label{sec:strategy_relation}
\label{sec:synergyproof}
From the definitions of the three different attacks and the strategies of how to prevent them, it becomes apparent that there are synergies among the objectives and some strategies entail others. More concretely, preventing value privacy attacks also prevents attacks on relationship anonymity and wormhole attacks. Furthermore, following our fee optimization algorithm also prevents all three security and privacy attacks (see \cref{sec:heuristics}). %

On a high level, for a given path $p$, let $V(p), R(p), W(p), F(p)$ be one out of potentially multiple sets of nodes that are at least required to be bypassed for preventing value privacy attacks, relationship anonymity attacks, and wormhole attacks, as well as for optimizing the fees, respectively. That is, if these nodes or a superset of them are bypassed, the corresponding attack is prevented. We define these sets formally 
as the return values of \cref{alg:bypassVP,alg:bypassRA,alg:bypassWH,alg:bypassFees} in \cref{sec:algorithms}.
Since $V(p), F(p)$ contain (at least) all adversarial nodes, optimizing for these objectives also prevents any other attack relying on on-path adversaries, such as denial-of-service attacks~\cite{aft20}.

\begin{theorem}
\label{thm:strategyRelation}
For any path p, $F(p) \supseteq V(p) \supseteq R(p)$ and $F(p) \supseteq V(p) \supseteq W(p)$.
\end{theorem}

\noindent\textit{Proof.} From the definitions of \cref{alg:bypassVP,alg:bypassRA,alg:bypassWH,alg:bypassFees}, we observe the following. $V(p)$ is the set containing all corrupted nodes on the path $p$. Intuitively, if it is not, then there is a corrupted node left on the path which is not bypassed, thus value privacy attacks are not prevented.
The two sets $R(p)$ and $W(p)$ do not have any honest nodes by definition (honest nodes do not need to be bypassed). $R(p)$ and $W(p)$ contain thus only malicious nodes, but they do not contain all the malicious nodes of path $p$. Consider for example path $s-c_1-h_1-c_2-h_2-c_3-r$ ($c_i$ representing corrupted and $h_i$ honest nodes), where $c_2$ is in $V(p)$, but not in $R(p)$.
It follows that $V(p) \supseteq R(p)$ and $ V(p) \supseteq W(p)$.

It remains to show that $F(p) \supseteq V(p)$. For this, we merely observe that the set $F(p)$ contains every node on the path $p$, which includes every corrupted node, which is $V(p)$.\hfill\qed

\section{Empirical evaluation}
\label{sec:heuristics}

We conducted extensive simulations to shed light on the optimized deployment of virtual channels, as well as to study the performance of our greedy algorithm (Section \ref{sec:algsandanalysis}).

\subsection{Input data preparation and methodology}
\label{sec:inputdata}
\label{sec:methodology}

\paragraph{Graph model and data} Recalling our model in~\cref{sec:model}, let $\GG := (\VV, \EE := \EEpay \cup \EEvirtual)$ be our VPCN graph with $\VV$ the set of nodes,
$\EEpay$ the set of PCs and $\EEvirtual$ (initially $\emptyset$) the set of PCs. %
We conduct our experiments on a snapshot of the Lightning Network (LN) from 
March 4, 2021~\cite{fiatjaf}.
The (largest connected component of the) graph contains 33k channels and 8k nodes that are part of at least one channel. For each channel, we read the capacity, the base, and the relative fee. The total network capacity is 1,167.4 BTC, the average base fee is 3,165 msat (millisatoshi), the average relative fee rate is 32,417 millionth of the satoshis transferred (one BTC is 100M satoshis). %
Due to the nature of PCNs the individual balance of each user remains private to an outsider. This is a common limitation for works investigating PCNs. We assume that each channel capacity is initially evenly distributed between both nodes.

\paragraph{Payments} For payments we sample $\rPay = 100$ random sender-receiver pairs in the graph and uniform payment amounts $\val \in [1,10]$ satoshis, modeling a micro-payment setting as the average channel capacity is 2.6M satoshis.

\paragraph{Constructing VCs}
We create VCs on top of the PCN, both for direct payments between endpoints and for routing other payments through the VCs. Users can charge fees (i) for establishing the VC if they are intermediaries or (ii) for routing payments through the VC if they are endpoints. 
\lukas{There exists no fee model for VCs in practice. Therefore, we interpret the base fee as what a hop charges for actively participating in the protocol, and the fee rate as what a hop charges for locking up $\amount$ coins, i.e., the opportunity cost of that node. We model the establishment fee of a VC with capacity $\amount$ to be the same as routing a payment of $\amount$ coins via that path (see \cref{sec:model}). Our solution is modular, other fee models can be used, and we discuss other possible fee models in 
\iftechreport
\cref{sec:discussion}.
\else
\cite[Section VII]{threebirds-IFIP-TR}.
\fi
If a VC is established, its routing fee is set to the fee of the initiating endpoint's underlying channel.}

\paragraph{Corruption model} 
Honest nodes assume there is an attacker who has an estimated budget and who corrupts nodes according to an estimated strategy. In order to corrupt a node $v \in \VV$, we say an attacker needs to spend the money that this node $v \in \VV$ has locked up in its neighboring channels, i.e., $\capacityLocked(v) := \sum_{w\in\VV \setminus\set{v}}(\pcid{v}{w}. \bal_1)$. 

The assumption that an attacker corrupts nodes that are most beneficial to it, while being cheap to place, is based on previous works \cite{tikhomirov2020quantitative,kapposEmpiricalAnalysisPrivacy2021,DBLP:journals/corr/abs-2103-08576,10.1145/3465481.3465761}. We parameterize the estimated adversary budget $\advBudget$ as a percentage of the total capacity in all edges of $\GG$. The absolute adversary capacity is $adv\_capacity := \advBudget \cdot \allowbreak \totalNetworkCapacity$. To choose the best-placed nodes, the adversary computes random payment paths and selects those nodes that appear most often on these paths. Note that no payment is actually carried out, only the paths are computed to find the most used nodes. Let $P$ be a list of $\numPay$ (e.g., 500) randomly chosen payment paths (i.e., paths of connected edges) in $\GG$. For every node $v\in \VV$, we let $\occ(v)$ be the number of their occurrence in $P$. We define the following cost-benefit ratio for every node $v$ as follows: 
$\costBenefit(v) := \frac{{\occ(v)}/{\numPay}}{{\capacityLocked(v)}/{adv\_capacity}}$.

Let $l$ be a list of every node $v\in \VV$ sorted by their cost-benefit ratio in descending order. We determine the list of all corrupted nodes $C$ iterating over $l$ and adding those for which the following condition holds after adding them: $\sum_{n\in C}(\capacityLocked(n)) \allowbreak \leq \advBudget$

\paragraph{Repeating payments needed for VCs to be cost efficient} If a VC is used only once, it will never cost fewer fees than routing a payment directly through the underlying PCs. 
Therefore, we investigate the effect of conducting our $\rPay$ payments multiple times. %

\paragraph{Measuring fees, security and privacy}
The cost of routing the $\rPay$ payments through the PCN without PCs is denoted as $\routePcn$. The cost of establishing the PCs to prevent a certain type of attack is denoted as $\establishVc$. The routing cost when using the PCs is $\routeVc$. We are interested in how the following ratio progresses as we increase the number of times that payments are repeated: $\feeRatio := \frac{\establishVc+\routeVc}{\routePcn}$.
We further measure how many payment paths are prone to a certain attack, with and without the VCs.

\subsection{Results}

\begin{figure*}[t]
    \centering
    \begin{minipage}{.6\columnwidth}
        \resizebox{\textwidth}{!}{
\begin{tikzpicture}
\tikzstyle{every node}=[font=\huge]
\pgfplotsset{compat=1.12}
\begin{axis}[
    enlarge x limits=0.15,
    enlarge y limits={0.15,upper},
    legend style={at={(0.5,-0.30)},
      anchor=north,legend columns=-1},
    ylabel={cost ratio VC/PCN},
    xlabel={how often payments are executed},
    minor xtick={5, 15, ..., 45},
    xtick={1,10,20,30,40,50},
    ymin=0.63,
    ymax=1,
    ]
\addplot coordinates {(1,0.984586542)(2,0.913758878)(3,0.895378097)(4,0.880735804)(5,0.871916124)(7,0.873312269)(10,0.864917125)(15,0.863964753)(20,0.860497995)(25,0.863538824)(30,0.854644703)(35,0.850418317)(40,0.860562582)(45,0.854118239)(50,0.853635465)}; %
\addplot coordinates {(1,0.981110741)(2,0.864720923)(3,0.832305287)(4,0.812264892)(5,0.800270438)(7,0.793828278)(10,0.785227544)(15,0.778175722)(20,0.782387932)(25,0.779255611)(30,0.768686823)(35,0.762099969)(40,0.769020854)(45,0.762372308)(50,0.769277612)}; %
\addplot [
    mark=triangle*,
    mark options={fill=yellow},
    line width=0.7pt,
    black, 
    mark size=3.5pt,
] coordinates {(1,0.978852965)(2,0.82293101)(3,0.76816162)(4,0.750868512)(5,0.735536173)(7,0.716248477)(10,0.701194126)(15,0.691864752)(20,0.69814495)(25,0.700640441)(30,0.684601035)(35,0.675082958)(40,0.681045468)(45,0.673366262)(50,0.67901254)}; %
\legend{budget:0.01,0.02,0.05}
\end{axis}
\end{tikzpicture}
}
    \end{minipage}%
    \begin{minipage}{.25\columnwidth}
        ~
    \end{minipage}
    \begin{minipage}{.6\columnwidth}
        \resizebox{\textwidth}{!}{
\begin{tikzpicture}
\tikzstyle{every node}=[font=\huge]
\pgfplotsset{compat=1.12}
\begin{axis}[
    enlarge x limits=0.15,
    enlarge y limits={0.15,upper},
    legend style={at={(0.5,-0.30)},
      anchor=north,legend columns=-1},
    ylabel={cost ratio VC/PCN},
    xlabel={how often payments are executed},
    minor xtick={5, 15, ..., 45},
    xtick={1,10,20,30,40,50},
    ymin=0.86,
    ymax=1,
    ]

\addplot coordinates {(1,1.000006707)(2,0.988558862)(3,0.984683682)(4,0.980098559)(5,0.98234015)(7,0.980750931)(10,0.980586029)(15,0.980274188)(20,0.982857301)(25,0.975080648)(30,0.97929191)(35,0.978854944)(40,0.978826154)(45,0.977236471)(50,0.979983142)}; %
\addplot coordinates {(1,0.999999619)(2,0.969647714)(3,0.962819036)(4,0.957891979)(5,0.952286612)(7,0.950513053)(10,0.950268517)(15,0.950884599)(20,0.951595142)(25,0.947425929)(30,0.94748397)(35,0.94495148)(40,0.943809292)(45,0.945208097)(50,0.950227448)}; %
\addplot [
    mark=triangle*,
    mark options={fill=yellow},
    line width=0.7pt,
    black, 
    mark size=3.5pt,
] coordinates {(1,0.999891728)(2,0.942237395)(3,0.918025986)(4,0.907373479)(5,0.899225644)(7,0.906508012)(10,0.894837515)(15,0.891165554)(20,0.885567423)(25,0.886600118)(30,0.885178264)(35,0.884354165)(40,0.889399804)(45,0.883104789)(50,0.88305959)}; %
\legend{budget:0.01,0.02,0.05}
\end{axis}
\end{tikzpicture}
}
    \end{minipage}
    \vspace{-1mm}
    {\color{lightgray}\hrule}
    \vspace{3mm}
    \begin{minipage}{.6\columnwidth}
        \resizebox{\textwidth}{!}{
\begin{tikzpicture}
\tikzstyle{every node}=[font=\huge]
\pgfplotsset{compat=1.12}
\begin{axis}[
    enlarge x limits=0.15,
    enlarge y limits={0.15,upper},
    legend style={at={(0.5,-0.30)},
      anchor=north,legend columns=-1},
    ylabel={cost ratio VC/PCN},
    xlabel={how often payments are executed},
    ymin=0.93,
    ymax=1,
    minor xtick={5, 15, ..., 45},
    xtick={1,10,20,30,40,50},
    ]
\addplot coordinates {(1,0.999611252)(2,0.978254189)(3,0.97571211)(4,0.96768685)(5,0.963902802)(7,0.967417428)(10,0.964117313)(15,0.96310272)(20,0.965296806)(25,0.966646306)(30,0.961218138)(35,0.95997979)(40,0.962160145)(45,0.960858332)(50,0.96313624)}; %
\addplot coordinates {(1,0.998914018)(2,0.973613922)(3,0.964448925)(4,0.960332236)(5,0.957723273)(7,0.957507185)(10,0.952337957)(15,0.950155243)(20,0.952530139)(25,0.950037774)(30,0.950266139)(35,0.952005798)(40,0.944351977)(45,0.947490408)(50,0.945892055)}; %
\addplot [
    mark=triangle*,
    mark options={fill=yellow},
    line width=0.7pt,
    black, 
    mark size=3.5pt,
] coordinates {(1,0.999186419)(2,0.972141604)(3,0.961414527)(4,0.95723531)(5,0.953695274)(7,0.947261613)(10,0.949124174)(15,0.946679003)(20,0.942891625)(25,0.942759498)(30,0.942912594)(35,0.945498188)(40,0.943383165)(45,0.946268447)(50,0.949548737)}; %
\legend{budget:0.01,0.02,0.05}
\end{axis}
\end{tikzpicture}
}
    \vspace{-0.4cm}
    \end{minipage}%
    \begin{minipage}{.25\columnwidth}
        ~
    \end{minipage}
    \begin{minipage}{.6\columnwidth}
        \resizebox{\textwidth}{!}{
\begin{tikzpicture}
\tikzstyle{every node}=[font=\huge]
\pgfplotsset{compat=1.12}
\begin{axis}[
    ylabel={cost ratio VC/PCN},
    xlabel={how often payments are executed},
    xmin=0, xmax=50,
    ymin=0, ymax=1.1,
    ytick={0,0.2,0.4,0.6,0.8,1},
    minor xtick={5, 15, ..., 45},
    xtick={1,10,20,30,40,50},
    legend pos=south east,
    ymajorgrids=true,
    grid style=dashed,
]

\addplot[
    color=blue,
    mark=square,
    ]
    coordinates {
    (1,1)(2,0.509088581)(3,0.341092643)(4,0.26523435)(5,0.215094584)(7,0.154139111)(10,0.118138035)(15,0.073030185)(20,0.06728271)(25,0.052382725)(30,0.047206373)(35,0.043690951)(40,0.049447057)(45,0.036022905)(50,0.034104231)
    };
    
\end{axis}
\end{tikzpicture}
}
    \vspace{-0.4cm}
    \end{minipage}
    \caption{Optimizing for value privacy (top left), relationship anonymity (top right), wormhole attack (bottom left), and fees (bottom right)}
    \label{fig:results}
    \vspace{-0.2cm}
\end{figure*}

We first study the effect that opening VCs while optimizing for each individual goal has on the other goals and on the fees. 
We fix an adversary budget and corrupt the nodes according to our corruption model. For each payment, we then use VCs (i) to optimize for security or privacy by preventing one of these attacks completely if that payment path is prone to that attack or (ii) to optimize for fees, both according to the algorithms outlined in \cref{sec:greedy}. Finally, we measure the impact this has on the two other attacks as well as on the fees. 
The full algorithm pseudocode can be found 
\iftechreport
in \cref{alg:bypassVP,alg:bypassRA,alg:bypassWH,alg:bypassFees} in \cref{sec:algorithms}.
\else
in~\cite[Section V-C]{threebirds-IFIP-TR}.
\fi

In our experiment, we investigate value privacy, relationship anonymity, and wormhole attacks. %
For these experiments we need to choose an adversary budget that results in meaningful security threats from all these attacks. By meaningful we mean that some of our paths (not 0 and not all of them) are susceptible to each of the three different attacks. For this, we need to compute the percentage of paths that are prone to which attack for different adversary budgets. We expand on this %
\iftechreport
in \cref{sec:prelimEval} 
\else
in~\cite[Section VI-C]{threebirds-IFIP-TR}
\fi
and end up choosing 1, 2 and 5\%.

\smallskip\noindent\textbf{Q1: How does preventing one attack affect the money spent on fees?}
We first measure the cost of routing payments through the PCN without VCs as a baseline. Then, we construct VCs, optimizing for value privacy, relationship anonymity, and the wormhole attack. 
After constructing the VCs, we measure the cost again. We measure the ratio according to our definition in \cref{sec:methodology}. The VCs are constructed according to the corrupted nodes on the payment paths. Since these paths are randomly chosen and thus different for every run, we conduct each experiment 100 times and compute the average, with the results 
shown in \cref{fig:results}.
We observe that for all three budgets, the cost ratio starts out around 1 for one payment. As the number of repeating payments goes up, the cost ratio decreases because the VCs are more effective. Additionally, the more nodes are corrupted and need to be bypassed, the more VCs are constructed and the better this ratio becomes. For relationship anonymity, this ratio goes down to 0.88, for wormhole attack to 0.95, for value privacy to 0.68.%

We observe that, generally speaking, bypassing nodes to prevent each attack has a positive effect on the fee ratio, if the payments are repeated more than once. The best effect can be seen in preventing value privacy attacks. Also, the ratio goes down more steeply for the first 10 payments, afterwards the effect is more flat.

\smallskip\noindent\textbf{Q2: How does optimizing for fees affect the money spent on fees?} %
Similar to when optimizing for security and privacy goals, we observe a steep decline in the ratio of money spent for fees when constructing VCs to the money spent for fees if we do not construct VCs. The decline slows down later.
This ratio halves if there are 2 sequential payments and continues to drop to 0.04 for 50, after which it is almost flat.

\begin{figure}
\centering
\caption{How many paths are prone to different attacks when optimizing for different goals for an adversary budget of 0.05.}
\label{tab:optimization}
\resizebox{\columnwidth}{!}{
\begin{tabular}{ccc|c|c|c}
                               &                                                  &               & \multicolumn{3}{c}{Paths prone to (PCN; VC)} \\ \hline
Optimizing                          & \# VC              & Avg VC length & VP attacks    & RA atk.    & WH atk.    \\ \hline
VP   & 126 & 3.1           & 97; 0         & 21; 0         & 35; 0         \\
RA                  & 21                          & 3.6           & 97; 84        & 21; 0         & 35; 28        \\
WH & 33  & 2.2           & 97; 97        & 21; 13        & 35; 0         \\
Fees                                    & 100                         & 5.6           & 97; 0         & 21; 0         & 35; 0        
\end{tabular}

}
\end{figure}
\smallskip\noindent\textbf{Q3: How do the different optimization strategies affect security and privacy?}
We already compared the effect of the strategies optimizing the different goals on the fees. Now we want to evaluate the effect that they have on the security and privacy goals. 
For this, we measure how many of our payment paths are prone to each of the different attacks. Then we construct the VCs optimizing each goal and measure how many paths are prone then.
In \cref{tab:optimization} we show for each optimization strategy, (i) how many VCs are constructed, (ii) the average length of each VC, and (iii) for each attack type two values $x;y$, where $x$ is the percentage of paths prone to the attack before building VCs and $y$ is the percentage of paths prone to the attack after building VCs. We notice that optimizing for value privacy also prevents the attacks on relationship anonymity and the wormhole attack. Furthermore, optimizing for fees prevents all three attacks we investigate. These results are in line with 
\cref{sec:strategy_relation}.

\subsection{Computing the percentage of prone paths for different adversary budgets}
\label{sec:prelimEval}

\begin{figure}[h]
\centering
\resizebox{0.7\columnwidth}{!}{
\begin{tikzpicture}
\begin{axis}[
    enlarge x limits=0.15,
    enlarge y limits={0.15,upper},
    legend style={at={(0.5,-0.20)},
      anchor=north,legend columns=-1},
    ylabel={\% of paths prone to each attack},
    xlabel={Adversary budget in \%},
    symbolic x coords={0.01,0.1,0.2,1,2,5,10,20,40,80},
    xtick=data,
    ]
\addplot [
    mark=square*,
    mark options={fill=orange},
    line width=0.7pt,
    black, 
    mark size=2.5pt,
] coordinates {(0.01,8.8)(0.1,52.8)(0.2,68.8)(1,85.4)(2,90.4)(5,96.9)(10,97.2)(20,98)(40,99.6)(80,99.8)};
\addplot [
    mark=triangle*,
    mark options={fill=purple},
    line width=0.7pt,
    black, 
    mark size=3.5pt,
] coordinates {(0.01,0)(0.1,0)(0.2,0)(1,3.8)(2,8.8)(5,20.9)(10,27.8)(20,46.2)(40,66.6)(80,72.8)};
\addplot [
    mark=diamond*,
    mark options={fill=green},
    line width=0.7pt,
    black, 
    mark size=3.5pt,
] coordinates {(0.01,1)(0.1,6.6)(0.2,14.2)(1,28.6)(2,30.4)(5,34.6)(10,23)(20,9.6)(40,1)(80,0)};
\legend{Attack:VP,RA,WH}
\end{axis}
\end{tikzpicture}
}
\caption{Number of prone paths}
\label{fig:paths_prone}
\end{figure}

We choose different adversary budgets ranging from 0.01\% to 80\% of the PCN capacity and check, how many of our payments are prone to the different attacks. We plot our results in \cref{fig:paths_prone}. We observe that value privacy attack is the cheapest, only $0.1\%$ yields more than half of the payments being prone. Note that an adversary budget of $0.1\%$ of the total capacity is still significant and requires in our data the staking of roughly 1.2 BTC. Achieving the wormhole attack is more expensive and the effectiveness peaks at around $5\%$ (58.4 BTC in our data) as adversary budget, before going down again. This is due to the fact that this attack requires honest nodes in between. The most expensive attack to mount for our corruption strategy is relationship anonymity. Only after $10\%$ adversary budget it affects more paths than wormhole attack. 

Note that an adversary can use its corrupted nodes to carry out any or all of these attacks. Note that there might exist other strategies for corrupting nodes that are more effective for relationship anonymity and wormhole attacks (cf.~\cite{tikhomirov2020quantitative}). %
From these results we choose $1$, $2$ and $5$ percent as the adversary budget we want to investigate further, as with these budgets there are some (and not all) paths that are prone to each of the attacks and to be able to compare them more easily.

\section{Discussion}
\label{sec:discussion}

\paragraph{Fair establishment fee model for intermediaries}
VCs are not used yet in practice, and thus, we do not know how fees are going to be charged.  In Section \ref{sec:heuristics}, we assumed that the fee of opening a VC with some capacity $\amount$ is the same as routing the payment of the same value through this path. Of course, other establishment fee models are possible.

First, in order to be fair to both the endpoints and intermediaries, VCs should have a limited lifespan. Remember that VCs require intermediaries to lock up some funds that they cannot use for other payments for which they would otherwise be able to charge fees. Note that this is different to PCs, where it is not problematic for two users to lock up their funds potentially indefinitely, as they do it only for their own funds. In other words, the fees that the intermediaries receive have to be proportionate to the time for which the money is locked up, i.e., the lifespan of a VC. Since channel capacities are finite, this implies that the lifespan is also finite.

The way we previously modeled fees is that VC creation fees are independent of the lifespan of a channel. In fact, one could argue that the fees for routing payments should then also be dependent on the collateral timeout, which, in practice, it is not. However, it is hard to model this in our experiments, as due to the nature of the PCN, it is unknown how many payments are processed in which time frames. For this reason, we chose our simplified fee model. 

In order for VCs to still be profitable in this fairer fee model, one would have to find a number $k$ of payment repetitions (for each of our $n$ payments), a lifetime of the VC $t$, the base fee $\fee$, the fee $\feeprop$ proportional to the amount and a fee $v$ that is proportional to the lifetime and the capacity of the VC, such that the inequality $\sum_{n}(\sum_{k}(\fee+\amount\cdot \feeprop)) \geq \sum_{n}(\fee + k\cdot\amount \cdot \feeprop + k\cdot\amount \cdot v\cdot t)$ holds.

This simplifies to the following: $(k-1) \cdot \fee \geq k \cdot \amount \cdot v\cdot t$. We note that this inequality is not exact, as the forwarded amount gets smaller when intermediaries already deduct fees. Nonetheless, the investigation of this or other fee models is interesting future work.

\paragraph{Utilizing bidirectionality of VCs} The greedy algorithms create virtual channels with transactions in one direction in mind. This means that on a path where several transactions are executed, the capacity is chosen as the sum of the amount of these transactions. In reality, the capacity can be lower if there are transactions in the other direction in the same time frame. Since we do not model any timing as mentioned above, we also do not capture this.

A practical scenario where this is useful might be where two payment providers route a substantial amount of payments through a hub back and forth. Note that the sum of all their payments can be very large, but since they send it back and forth the capacity of the virtual channel can be smaller.

\paragraph{Adversaries agreeing to open VCs} One could argue that an adversary trying to conduct an on-path attack would not agree to create a VC because the adversary would hinder its own attack capabilities. However, refusal to participate in creating a VC will be noticed by the sender. Thus, a sender concerned with security and privacy can always prevent such an attack by finding an alternative path.

\paragraph{Blocking capacity with VC} If a VC is used only sparsely, then it might block the capacity of the underlying payment channels. Thus, it should either be used for routing many payments or be open only for a short amount of time.

\paragraph{VC routing fee} Similar to the establishment fee, it is unclear what fee the users of a virtual channel will charge in practice for letting other users route their payments through the virtual channel. In this work, we assume this fee to be the same as fees that are charged for one of the underlying channels.

\paragraph{Privacy of VC creation} Following our strategies, we aim to prevent the on-chain privacy attacks on value privacy and relationship anonymity for payments. However, since opening or closing a virtual channel is an operation in which potentially corrupted intermediaries participate, some information might be leaked while opening or closing a virtual channel.

We assume that VCs are used in the same way as payment channels, e.g., for routing payments through them. To make other users aware of accepting payment being routed through channels, channels need to be announced publicly. Therefore, we disregard privacy concerns for VC opening/closing and instead assume that endpoints and capacity of VCs are announced publicly, thereby already leaking the information on endpoints and capacity.

\paragraph{Using VCs for routing} In our evaluation of the greedy algorithm, we consider the payments that are repeated once or multiple times and conclude that if payments are conducted more than once along a path, opening a direct VC is the best strategy. However, for payments that are conducted only once, the VCs that were created in this fashion can be used for having shorter paths, which means having fewer fees and being less at risk for attacks due to having fewer intermediaries.

\section{Related work}
\label{sec:relwork}

Over the last years, significant research efforts have been devoted
to the design and analysis of efficient and secure payment channel networks~\cite{neudecker2018network,gudgeon2020sok,csur21crypto}.
Motivated by topology-based attacks~\cite{rohrer2019discharged}, the possibility of route hijacking~\cite{aft20}
as well as vulnerabilities, e.g., related to the privacy~\cite{tang2020privacy,icissp20,kapposEmpiricalAnalysisPrivacy2021} and anonymity 
of PCN users~\cite{Roher2020Counting,romitiCrossLayerDeanonymizationMethods2021}, to just name a few examples, 
much existing literature revolves around network connectivity \cite{rohrer2019discharged}, the payment routing system \cite{aft20,malavolta2017concurrency}, 
as well as privacy aspects, e.g., of route discovery~\cite{ifip21pir,tang2020privacy}.

To ensure anonymity, payment-channel networks usually rely on privacy-enhancing cryptographic schemes 
(e.g., onion routing) to implement 
the 2-phase commit payment operation. 
PrivPay~\cite{moreno-sanchezPrivacyPreservingPayments2015},  SilentWhispers~\cite{silentwhispers}, Fulgor/Rayo~\cite{malavolta2017concurrency}, AMHL~\cite{malavoltaAnonymousMultiHopLocks2019} provide 
privacy-preserving multi-hop payment protocols which come with formal guarantees.
SpeedyMurmurs~\cite{roos2017settling} formalizes and addresses
concrete notions of privacy in the context of payment routing. 
SpiderNetwork~\cite{sivaraman2018routing} improves the effectiveness of source
routing in a dynamic PCN by favoring routes that minimize the balance difference using on-chain
rebalancing. 
A privacy-preserving approach to discovering low-cost routes was recently presented
by Pietrzak et al.~\cite{ifip21pir}.
Blitz~\cite{aumayrBlitzSecureMultiHop2021} is a 1-phase payment scheme, which, similar to AMHL, provides security against wormhole attacks~\cite{malavoltaAnonymousMultiHopLocks2019}. None of the payment-based approaches, however, hide the value of the payment to intermediaries or decrease routing fees.

Therefore, an intriguing approach to improving the security and efficiency of payment channel networks is the use of virtual channels. These have been introduced by Dziembowski et al.~\cite{dziembowski2017perun} to overcome the requirement that intermediaries along a channel route need 
to be online (a concern also considered in \cite{mccorry2019pisa,avarikioti2019brick}) 
and explicitly confirm all mediated transactions. Recent work has extended the deployment scope of virtual channels, introducing efficient
protocols that are compatible with Bitcoin and other popular cryptocurrencies~\cite{aumayrBitcoinCompatibleVirtualChannels,DBLP:conf/cans/JourenkoLT20,cryptoeprint:2021:855}.

While existing literature on virtual channels revolved around protocol design aspects, to the best of our knowledge, our paper is the first to investigate the problem of optimizing the allocation of virtual channels in order to improve the security and efficiency of PCNs. 
In parallel work~\cite{khamis2021demand}, Khamis and Rottenstreich 
studied how to amortize the creation of new channels through reduced routing costs,
however, without accounting for security aspects.

\section{Conclusion}\label{sec:conclusion}

Motivated by the potential benefits of virtual channels to reduce transaction fee costs as well as to improve security and privacy guarantees in PCNs, we presented a first systematic study of the virtual channel setup problem.
We have shown that the problem can be formulated as an optimization
problem and proved that the problem is NP-hard.
We presented a fast greedy algorithm and using simulations on the Lightning Network, we confirmed the benefits of our optimization approach.
We also modeled the VPCN cost optimization problem as an integer linear program (ILP) for obtaining exact solutions.
While solving the ILP is computationally expensive (this is expected due to the problem's hardness), we discussed methods for reducing its running time by reducing the solution quality.

We believe that our work opens several interesting avenues for future research, such as studying different fee models, the effect of timing, i.e., adding time frames in which transactions are to be executed, VC lifetimes, or more dynamic strategies for the attacker or how the attacker can best react to the countermeasures proposed in this work.

\paragraph{Acknowledgements}
This work has been supported by the European Research Council (ERC) under the Horizon 2020 research (grant 771527-BROWSEC); 
by the Austrian Science Fund (FWF) through the SpyCode SFB project F8510-N and the project CoRaF
(grant agreement 2020388); by CoBloX Labs; by the Austrian Federal Ministry for Digital and Economic Affairs, the National
Foundation for Research, Technology and Development and the Christian Doppler Research Association through the Christian Doppler Laboratory Blockchain Technologies for the Internet of Things (CDL-BOT); by the German Federal Ministry of Education and Research (BMBF) grant 16KISK020K (6G-RIC); 
by MCIN/AEI/10.13039/501100011033 and European Union NextGenerationEU/PRTR through PRODIGY Project (TED2021-132464B-I00), grant PID2022-142290OB-I00 and grant IJC2020-043391-I.

\bibliography{bibfileTR,libraryTR}

\end{document}